\documentclass[%
 amsmath,amssymb,
 aps,
prd,showpacs,onecolumn,superscriptaddress
]{revtex4-2} 

\usepackage{graphicx} % Include figure files
\usepackage{subfigure} 
\usepackage{dcolumn} % Align table columns on decimal point

\usepackage{amsmath} 
\usepackage{amssymb} 
\usepackage{bm} 
\usepackage{color} 

\def\blue#1{\textcolor{blue}{#1}}

\usepackage[normalem]{ulem} 
\usepackage[dvipsnames]{xcolor} 
\usepackage{hyperref} 
\usepackage{bm} % bold math
\usepackage{amsthm,amsmath,amssymb} 
\usepackage{mathrsfs} 
\usepackage{url}

\begin{document}

\title{Influence of mass-ratio corrections in extreme-mass-ratio inspirals for testing general relativity} 

\author{Ping Shen} 
 \affiliation{Shanghai Astronomical Observatory, Shanghai, 200030, China} 
\affiliation{School of Astronomy and Space Science, University of Chinese Academy of Sciences,
Beijing, 100049, China} 
\author{Wen-Biao Han} 
 \email{wbhan@shao.ac.cn} 
\affiliation{School of Fundamental Physics and Mathematical Sciences, Hangzhou Institute for Advanced Study, UCAS, Hangzhou 310024, China} 
\affiliation{Shanghai Astronomical Observatory, Shanghai, 200030, China} 
\affiliation{School of Astronomy and Space Science, University of Chinese Academy of Sciences,
Beijing, 100049, China} 
\affiliation{Shanghai Frontiers Science Center for  Gravitational Wave Detection, 800 Dongchuan Road, Shanghai 200240, China} 

\author{Chen Zhang} 
\affiliation{Shanghai Astronomical Observatory, Shanghai, 200030, China} %
\author{ Shu-Cheng Yang} 
 \affiliation{Shanghai Astronomical Observatory, Shanghai, 200030, China} 
\author{ Xing-Yu Zhong} 
 \affiliation{Shanghai Astronomical Observatory, Shanghai, 200030, China} 
\affiliation{School of Astronomy and Space Science, University of Chinese Academy of Sciences,
Beijing, 100049, China} 
\author{ Ye Jiang} 
 \affiliation{Shanghai Astronomical Observatory, Shanghai, 200030, China} 
\affiliation{School of Astronomy and Space Science, University of Chinese Academy of Sciences,
Beijing, 100049, China} 
\author{Qiuxin Cui} 
 \affiliation{Department of Astronomy, University of Science and Technology of China, Hefei 230026, China} 
\date{\today} % It is always \today, today,

\begin{abstract} 
The popular waveform templates of extreme-mass-ratio inspirals usually omit the mass-ratio corrections in the conservative dynamics, and employ adiabatic approximation to include the radiation reaction. With the help of effective-one-body dynamics, we investigate the influence of mass-ratio corrections  in the conservative part on the gravitational waves. We find that for the extra-relativistic orbits, the mass-ratio corrections can produce obvious orbital dephasing even for very small mass-ratio and then affect the waveforms. More importantly, it is noted that omitting the mass-ratio corrections in waveform templates may induce a fake signal of deviation from general relativity.  
\end{abstract} 

\maketitle
\section{\label{intro} Introduction} 

Extreme-mass-ratio-inspirals (EMRIs)    \cite{2007Intermediate,PhysRevD.95.103012}, consisting of   a stellar-mass compact object (SCO: white dwarf, neutron star or
black hole (BH) )  inspiraling into a supermassive black hole (SMBH), are the prime sources of the space-borne gravitational wave (GW)  detectors, such as the  Laser Interferometer Space Antenna (LISA)    \cite{Danzmann2017LISALI}, Taiji   \cite{Hu2017TheTP,zhong2023exploring}  and TianQin   \cite{Luo_2016}. 

EMRIs are very important in astrophysics, cosmology, and basic physics. In astrophysics, EMRIs will provide us with   relevant information on stellar dynamics in galactic nuclei, from which  we can infer the formation mechanism of EMRIs   \cite{2007Intermediate,king2005aligning,hughes2003black,rees2006massive}. In cosmology, the detection of a single EMRI event and its electromagnetic counterpart can provide an accurate measurement of the luminosity and the redshift  to estimate  the Hubble constant   \cite{wang2019bright,schutz1986determining,macleod2008precision,Zhao_2022}. For basic physics, EMRIs can be used to test general relativity (GR)  and the nature of BHs in the strong-field regime   \cite{gair2013testing,barack2007using}. 
As an extreme mass-ratio, SCO may be considered approximately as a test particle in the background of the central SMBH. 
The emitted GWs reflect the space-time information of the central object. If extracting this information from the gravitational-wave signals, we can accurately distinguish whether the central object is a Kerr BH or another corresponding object.

Extensive work has been done to quantify the ability of space-based GW detectors (LISA, Taiji, TianQin, etc)  to measure the deviation of multipole moments from the Kerr BH and then test GR \cite{PhysRevD.56.1845,PhysRevD.75.042003,PhysRevD.69.124022,Fransen_2022,PhysRevD.100.084055}.
In particular, all of the multipoles of Kerr BH are determined by its mass $M_0=M$ and spin $S_1=J=Ma$ :
\begin{equation} 
   M_l+iS_l=M (ia) ^l,% M_{2l} =M (-a^2) ^l, S_{2l+1} =Ma (-a^2) ^l
\end{equation} 
where $M_l$ and $S_l$ are the mass and mass-current multipole moments, respectively, and $a$ is the standard Kerr spin parameter.
For example, Ryan found that for a simplified case of circular, equatorial orbits, a LISA observation can measure the mass quadrupole moment to the accuracy of $\Delta M_2/M^3 \sim 0.0015-0.015$    \cite{PhysRevD.56.1845}. Barack and Cutler extended Rayan's work to generic orbits and considered the modulations caused by  satellite motions   \cite{PhysRevD.75.042003}.  Collins and  Hughes constructed the spacetime of bumpy BHs, which are similar to the Kerr BH, but with different multipoles. If the BH candidates are indeed Kerr BHs in GR, then their bumps should be zero   \cite{PhysRevD.69.124022}.
Furthermore, Kwinten Fransen  and Daniel R. Mayerson estimated the accuracy of  LISA EMRIs to measure  the  equatorial symmetry breaking using the lowest-lying odd-parity multipole moments $S_2$ and  $M_3$. They found  that  $S_2/M^3$ will typically be detectable with a measurement accuracy of $\Delta (S_2/M^3)  \approx 1\%$   \cite{Fransen_2022}.

These tests require an accurate waveform template.
With the accumulation of observation time from several months to a few years, the signal-to-noise ratio (SNR)  of EMRIs could become large enough to be detected by matched filtering    \cite{2007Intermediate}. However, the accuracy of the waveform template is also expected to be very high to detect such long-duration signals and extract the source parameters. For detection purposes, it is required that the dephasing over  the signal's duration should be less than a few radians. For source parameter extraction, the dephasing should satisfy  $\Delta \phi \lesssim \rm{SNR}$  \cite{PhysRevD.95.103012,PhysRevD.78.064028}. As in Ref.  \cite{Barack_2018}, the GW phase $\phi$ can be expanded as 
\begin{equation}
    \phi=\nu^{-1}\phi_0 + \nu^{0}\phi_1 +O(\nu)\label{eq:phiPA},
\end{equation}
where $\nu$ is the mass ratio. The first term is the ``adiabatic'' term, i.e. 0th post-adiabatic (0PA) term, which involves the dissipative part of the first-order self-force (1SF). The second term is the 1st post-adiabatic (1PA) term, involving the conservative part of the 1SF and the  dissipative part of the second-order self-force (2SF), which is required for exact EMRI models \cite{2018mgm..conf.1953H,wardell2023gravitational}. Models that get  $\phi_0$ might be enough to detect most signals, but models that get  both $\phi_0$ and $\phi_1$ should be enough for precise parameter extraction \cite{GSF-ppt}.

Currently, there is a relatively accurate method based on the black hole perturbation theory (BHPT), i.e. the waveform can be obtained by solving the Teukolsky equation   \cite{teukolsky1973perturbations}. This method is based on frequency domain decomposition and treats the SCO as a test particle. 
The ongoing gravitational self-force (GSF) program \cite{Barack_2018} is a specific expansion within BHPT that aims to generate EMRI waveforms satisfying the requirements of the EMRI science. In this approach, EMRI is treated as a point mass orbiting a black hole and the dynamics can be described by the equation of motion of the mass, including the influence of the interaction with the self-field, i.e. the GSF \cite{10.1093/ptep/ptv092}.
Among the GSF formalisms, the recent post-adiabatic (PA) waveforms \cite{wardell2023gravitational}  are the most accurate waveforms to date.
Some other methods, such as the kludge models   \cite{barack2004lisa,babak2007kludge},  adiabatic waveforms \cite{PhysRevLett.126.051102,PhysRevD.103.104014},  and the effective-one-body (EOB)  formalism   \cite{PhysRevLett.104.091102,PhysRevD.84.044014,Han_2017,2014IJMPD..2350064H,PhysRevD.83.044044}, make it possible to efficiently calculate a large number of relatively accurate waveforms. 

Nowadays, kludge waveforms have been greatly developed, like analytical kludge (AK)     \cite{barack2004lisa}, numerical kludge (NK)     \cite{babak2007kludge}, augmented analytical kludge (AAK)     \cite{chua2017augmented}, etc. 
These kludge models can generate waveforms more quickly than using the Teukolsky equation or the GSF method, and are often used for EMRI data analysis.
However, they are less accurate because they represent approximations to the 0PA (adiabatic)  waveforms  \cite{wardell2023gravitational}.
They regard SCOs as test particles in the orbital evolution using  the matched asymptotic expansions  \cite{PhysRevD.12.2183}, ignoring the mass-ratio corrections in the conservative dynamics, which is related to the conservative part of the 1SF. However, in the future, we can incorporate the 1PA term into these waveform templates to model EMRIs exactly. 
 While not being enough for parameter estimation, models like the AAK might be sufficient for detection and they have been heavily used for many LISA measurement studies.

Hughes \cite{PhysRevD.103.104014} et al. computed adiabatic (0PA) waveforms for EMRIs by “stitching” together a long inspiral waveform from a sequence of waveform snapshots, each of which corresponds to a particular geodesic orbit. But some effects are neglected by this adiabatic approximation, such as the conservative self-force and spin-curvature coupling. However, we can extend this framework to include these effects beyond the adiabatic approximation in the future.
Recently, Wardell  \cite{wardell2023gravitational} et al. produced the 1st post-adiabatic (1PA) waveforms for nonspinning compact binaries undergoing a quasicircular inspiral, which  will be invaluable to accurately model EMRIs for the LISA mission.

As in Eq. \eqref{eq:phiPA}, the 1PA contribution is nonnegligible for all mass ratios, and the 1PA term  is both necessary and sufficient for EMRI modelling \cite{Barack_2018,2018mgm..conf.1953H,wardell2023gravitational}. So the lack of 1PA term will  lead to inaccurate waveform templates,  which may give rise to some errors in  GW parameter estimation  \cite{arun2009massive,huerta2009influence}. Especially, using a waveform template that is not accurate enough can lead to the detection of a spurious signal deviating from the Kerr black hole case.
Based on a simple linearized analysis and Fisher matrix calculations, Moore  \cite{moore2021testing}  et al. estimated the waveform errors accumulated among  GW events. They found that the evidence for the deviation from GR grows as the catalog size increases. Qian Hu and John Veitch   \cite{Hu_2023} investigated the impacts of overlapping signals and inaccurate waveform models on tests of GR. 
They confirmed that systematic errors could accumulate when combining multiple GW events and could lead to a false deviation from GR in some cases.

In our work, we are interested in evaluating the impact of inaccurate waveforms due to the ignoring of mass-ratio corrections (associated with the 1PA term), which may mistakenly lead to fake deviations from GR in parameter estimation. In other words, even if GR is a correct description of nature, there is a risk of making a wrong judgment on the test of gravity theory if mass-ratio correction is ignored in waveform templates. 

In the previous papers, taking  advantage of the EOB formalism   \cite{PhysRevLett.104.091102,PhysRevD.84.044014,Han_2017,2014IJMPD..2350064H,PhysRevD.83.044044}, some of us discussed the influence of   mass-ratio  corrections  in conservative dynamics   \cite{zhang2021geometrized,Zhang_2021}.
In the present work, we make a further and detailed investigation of mass-ratio corrections on the orbital motion (orbit, frequency, and phase)  and gravitational radiation. We also show the discrepancy of the Teukolsky-based energy fluxes by using the EOB orbit and the test particle one. 
We generate our waveforms by the following scheme: First, we improve the hybrid scheme of fluxes  \cite{gair2006improved}  (hereafter GG) to obtain the orbital evolutions with dissipative 1PA term. And then we solve the EOB dynamical equations including mass-ratio corrections (conservative 1PA term) to get the orbital frequency. Finally, the waveform are computed via the Teukolsly formalism.
More importantly, based on the Bayesian analysis, we evaluate the possibility of inducing a fake signal of deviation from GR due to omitting the mass-ratio corrections in waveform templates. We find that for the case of mass-ratio $\nu \lesssim 10^{-5} $, the mismatch  between the EOB and test particle waveforms is so small  that can be ignored. However, for the case of $\nu \gtrsim 10^{-5} $,  the mismatch is much larger, and there is a risk of making the incorrect judgment that we have detected a deviation from GR. 

This paper is organized as follows. In section \ref{EOB}, the conservative dynamics part  of analytical EOB formalism which includes the first-order mass-ratio corrections is introduced in detail. 
In section \ref{waveforms}, considering the gravitational radiation, we introduce the Teukolsky equation and generate the Teukolsky-based waveforms of EMRI by combining the EOB orbit and the post-Newtonian (PN)  fluxes fitted by the Teukolsky one.
In section \ref{bayes}, we consider EMRI models with different mass-ratios and SNRs to illustrate the waveform mismatches using the EOB orbit and the test particle one. We also calculate their Bayesian factors in favor of a GR deviation.
Finally, in section \ref{conclusion},  we draw the conclusions and discuss our results. Throughout this paper, we set $G=c=1$, where $G$ is the gravitational constant and $c$ is the speed of light.

\section{\label{EOB} Effective-one-body dynamics} 

The basic idea of the EOB formalism \cite{damour2001coalescence,PhysRevD.59.084006} is to simplify the conservation dynamics of the two-body problem in GR to the geodesic motion of the test particle in the reduced space-time. That is to say, the two-body problem of the SMBH with a mass $m_1$ and the smaller compact object with a mass $m_2$ is transformed into the single-body problem of the  test particle with the reduced mass $\mu=m_1 m_2/ (m_1+m_2) $ moving in the equivalent external metric field. This is equivalent to the post-Newtonian expansion of the traditional two-body relative motion. The EOB Hamiltonian is written as follows
\begin{equation} 
    H_{\rm EOB} =M\sqrt{1+2\nu (\hat{H} _{\rm eff} -1) }  ,
\end{equation} 
where  $M=m_1+m_2$, $\nu=m_1 m_2/M^2=\mu/M$, and $\hat{H} _{\rm eff} =H_{\rm eff} /\mu$, $H_{\rm eff} $ is the effective Hamiltonian. %$H_{\rm eff} =H_{NS} +H_S-\frac{\nu } {2r^3} S^2_* \approx H_{NS} $
Once the expression of the Hamiltonian $H_{\rm EOB} $ is determined,  the equations of motion can be written as
\begin{equation} 
        \frac{d \bm r} {d t} =\frac{\partial H_{\rm EOB} } {\partial \bm P}, \    \frac{d \bm P} {d t} =-\frac{\partial H_{\rm EOB} } {\partial \bm r}  + \bm F,
\end{equation} 
where $\bm r$ represents the coordinate of the small compact object, and $\bm P$ is the momentum.
We use $\bm F$ to denote the radiation reaction force of GW, which  causes the energy and angular momentum of the two-body system to be no longer conserved. In the conservative dynamic part, GW  radiation is ignored, that is, $\bm F=0 $.

The effective Hamiltonian is given by   \cite{zhang2021geometrized} 
\begin{equation} 
    H_{\rm eff} =H_{\rm NS} +H_{\rm S} -\frac{\nu } {2r^3} S^2_* \approx H_{\rm NS}  ,
\end{equation} 
where $H_{\rm S} $ and $H_{\rm NS} $ represent the Hamiltonian of the particle with and without spin,  $S_*=aM (m_2/m_1) $ is the effective spin of the particle, and $a$ is the deformed-Kerr spin parameter. $H_{\rm S} $ and $S_*$ are small enough to be ignored within the range of error precision   \cite{barausse2011extending}. The deformed-Kerr metric  is given by   \cite{barausse2010improved} 
\begin{subequations} \label{eq:metric} 
    \begin{gather} 
     g^{tt} =-\frac{\Lambda_t} {\Delta_t \Sigma},\\
     g^{rr} =\frac{\Delta_r} { \Sigma},\\
     g^{\theta \theta} =\frac{1} { \Sigma},\\
     g^{\phi \phi} =\frac{1} {\Lambda_t}  (-\frac{{\Tilde{\omega} _{\rm fd} } ^2} {\Delta_t \Sigma} +\frac{\Sigma} {\sin^2{\theta} } ),\\
     g^{t \phi} =-\frac{\Tilde{\omega} _{\rm fd} } {\Delta_t \Sigma},
    \end{gather} 
\end{subequations} 
 with 
\begin{subequations} 
    \begin{gather} 
     \Lambda_t= (r^2+a^2) ^2-a^2\Delta_t \sin^2{\theta},\\   
     \Delta_t=r^2[A (u) +\frac{a^2} {M^2} u^2],\\
     \Sigma=r^2+a^2 \cos^2{\theta},\\
     \Delta_r=\Delta_t D^{-1}  (u),\\
     \Tilde{\omega} _{\rm fd} =2a M r+\omega^{\rm fd} _1 \nu \frac{a M^3} {r} +\omega^{\rm fd} _2 \nu \frac{ M a^3} {r},
    \end{gather} 
\end{subequations} 
where $\omega^{\rm fd} _1=-10$, $\omega^{\rm fd} _1=20$   \cite{2014IJMPD..2350064H,PhysRevD.78.044002,Barausse_2009}, $u=\frac{M}{r}$, $A$ and $D$  are metric potentials for the EOB formalism mentioned in Ref.   \cite{steinhoff2016dynamical}, which are given in Appendix A.
We still call the coordinate ($t, r, \theta, \phi$) used here the Boyer-Lindquist-type coordinate (i.e., in coordinates that reduce to Boyer-Lindquist coordinates if the quadrupole perturbation is zero, thus reducing the spacetime to pure Kerr)  \cite{PhysRevD.81.084024}.

As in Ref.   \cite{zhang2021geometrized}, the orbital evolution equations can be expressed as the function of $ (\xi, \chi, p, e, \iota) $ 
\begin{subequations} 
\begin{gather} 
\Dot{\xi} =-\frac{ (1+e\cos{\xi} ) ^2} {epM\sin{\xi} }   \frac{g^{rr} \hat{P} _r} {E/M (g^{tt}  \hat{H} _{\rm eff} -g^{t \phi}  \hat{L} _z)  }  ,\\
\Dot{\chi} =-\frac{g^{\theta \theta } \sqrt{ (a^2 (1-\hat{H} ^2_{\rm eff} )  (z^2_{+} -z^2_{-} \cos^2{\chi} ) } } {E/M (g^{tt}  \hat{H} _{\rm eff} -g^{t \phi}  \hat{L} _z)  }, \\
\Dot{\phi} =\frac{g^{t \phi} -[g^{tt} g^{\phi \phi} - (g^{t \phi} ) ^2] \frac{\hat{L} z} {g^{tt}  \hat{H} _{\rm eff} -g^{t \phi}  \hat{L} _z} } {g^{tt} E/M}, 
\end{gather} 
\end{subequations} 
where $\cos^2{\theta} =\cos^2{\theta_{\rm min} }  \cos^2{\chi}=\sin^2{\iota}  \cos^2{\chi}$, $ r=pM/ (1+e\cos{\xi})$, $p$ is the semilatus rectum, $e$ is the eccentricity of the orbit, and $\iota$ is the orbital inclination.
$\xi$ varies from $0$ to $2\pi$ corresponding to $r$ going through a complete cycle, and $\chi$ varies from $0$ to $2\pi$ corresponding to $\theta$ oscillating through its full range of motion  \cite{zhang2021geometrized}.
$L_z$ is the angular momentum in the z-direction, $P_\theta$, $P_r$, and $P_\phi$ are the polar, radial, and  azimuthal angular momentum, and $E$ is the total system energy. The expressions of these parameters are given in Appendix A (Eqs.  \eqref{eq:Lz}--\eqref{eq:E}, Eqs.  \eqref{eq:ptheta}--\eqref{eq:pphi}).

By solving the above ordinary differential equations by numerical integration, we can obtain 
$\xi $, $\chi $, $\phi $ in the coordinate time $ t $. Projecting the Boyer-Lindquist coordinate to the spherical coordinate grid,  the corresponding Cartesian coordinate system is defined by
\begin{subequations} 
\begin{gather} 
\Tilde{x} =\frac{p\cos{\phi} \sqrt{1-z^2_- \cos^2{\chi} } } {1+e\cos{\xi} }, \\
\Tilde{y} =\frac{p\sin{\phi} \sqrt{1-z^2_- \cos^2{\chi} } } {1+e\cos{\xi} },  \\
\Tilde{z} =\frac{p z_- \cos{\chi} } {1+e\cos{\xi} }  ,
\end{gather} 
\end{subequations} 
where $z_-=\cos{\theta_{\rm min} }$.
%轨道图
\begin{figure} 
    \centering
    \begin{subfigure}[]{
	\includegraphics[scale=0.6]{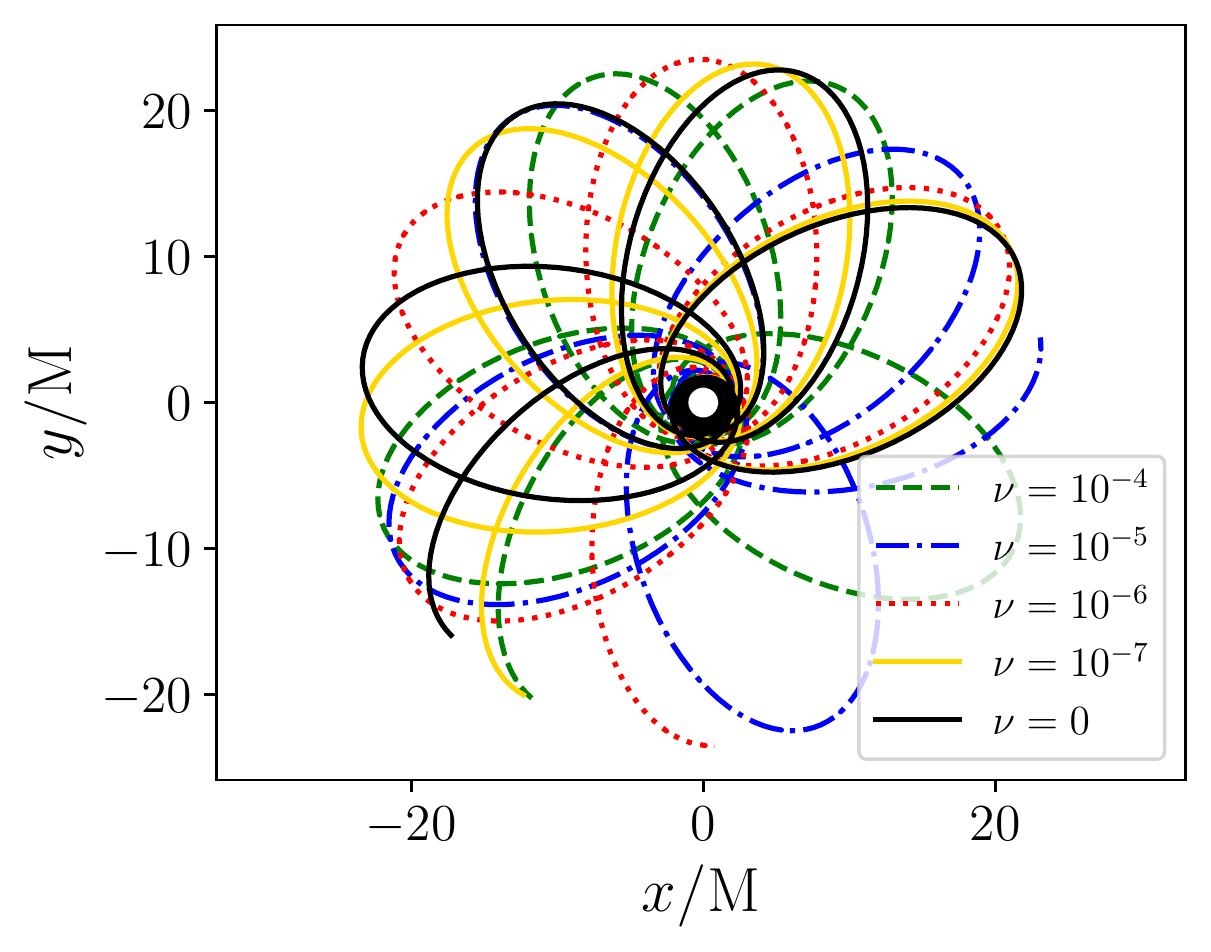}         
} 
	\end{subfigure} 
    \begin{subfigure}[]{
	\includegraphics[scale=0.6]{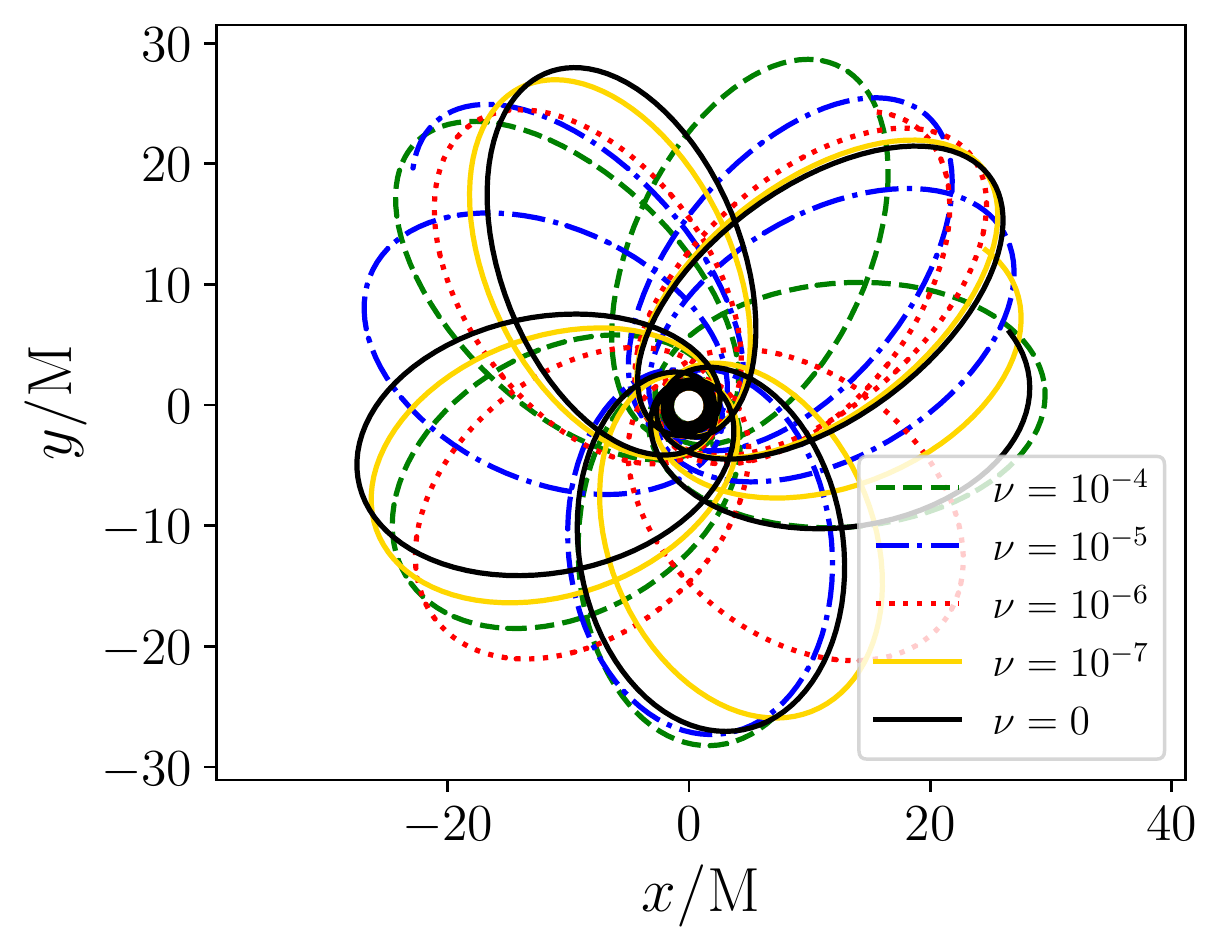}  
} 
        \end{subfigure} 
        
\begin{subfigure}[ ]{
	\includegraphics[scale=0.6]{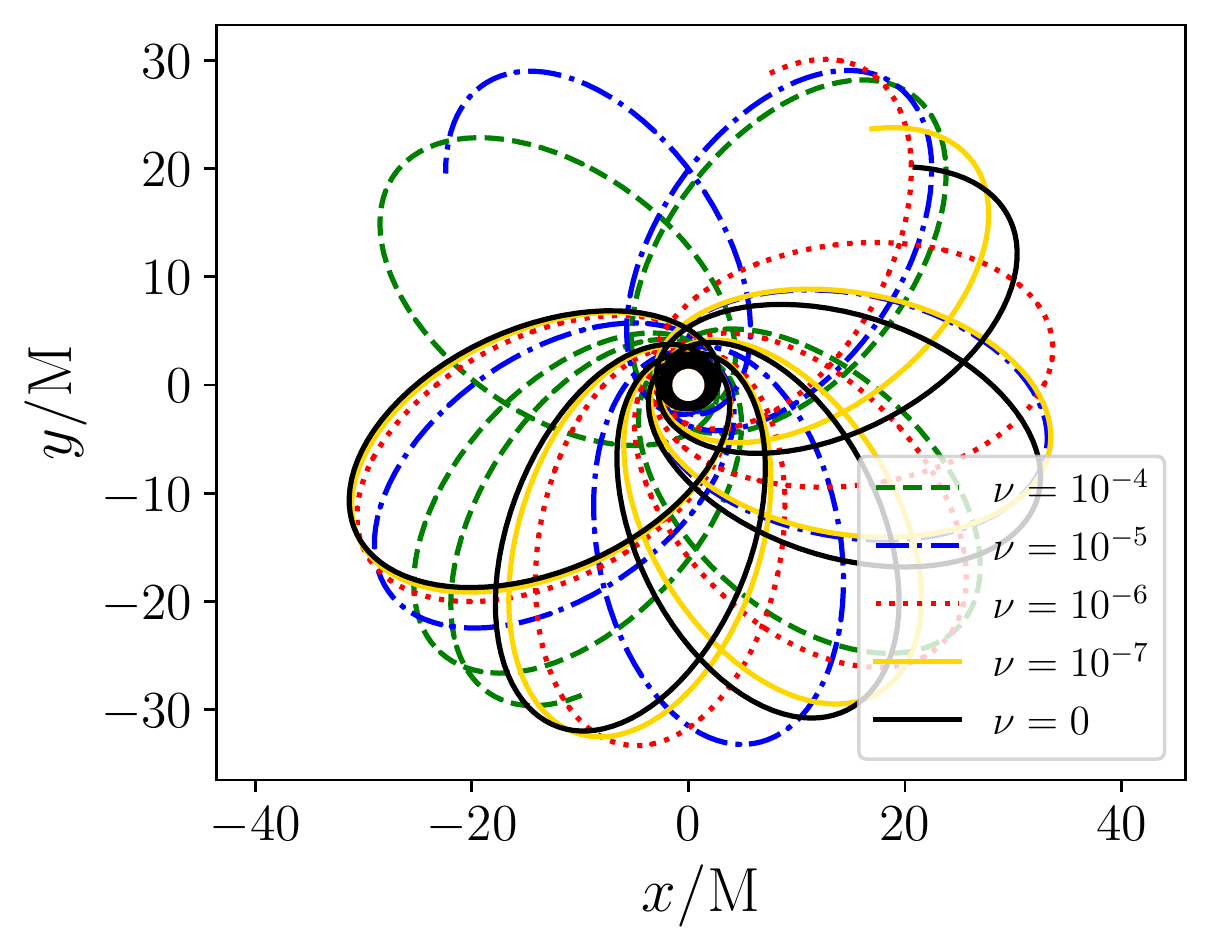} 
} 
	\end{subfigure}  	
 \begin{subfigure}[]{
	\includegraphics[scale=0.6]{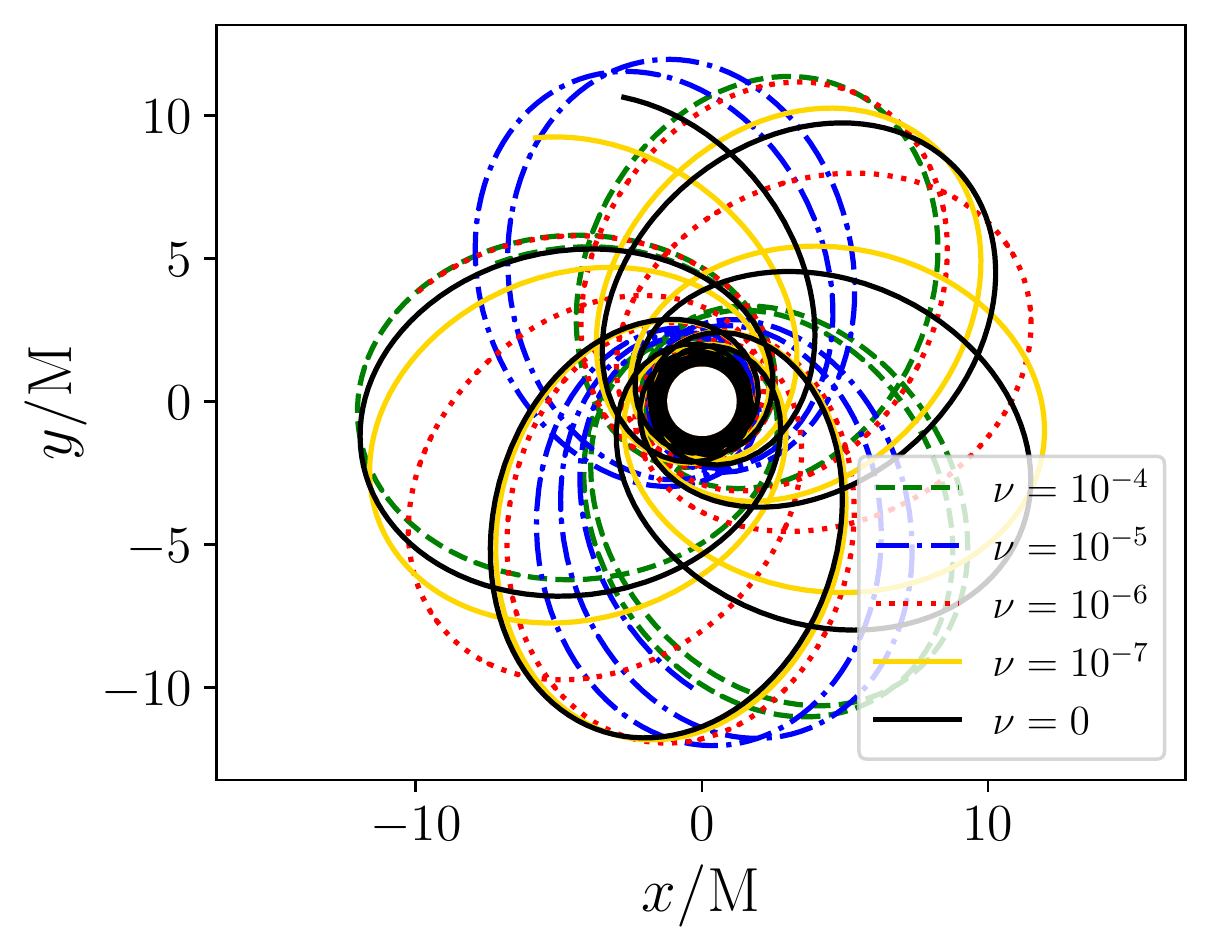}
 }
	\end{subfigure} 
    \caption{\label{fig:orbit}  Four orbits  in the x-y plane of different orbital parameters. The green, blue, red, yellow, and black lines  represent $\nu=10^{-4},10^{-5},10^{-6},10^{-7} $, and   $\nu=0$, respectively. Note that $\nu=0$ denotes the case of the test particle. (a): $ a=0.99,e=0.9,\iota=0^{\circ},p=2.35 M$, (b): $ a=0.95,e=0.9,\iota=30^{\circ},p=3.15 M$, (c): $a=0.9,e=0.9,\iota=10^{\circ},p=3.39 M$, (d): $a=0.98,e=0.8,\iota=0^{\circ},p=2.41 M$ } 
\end{figure}

In  Fig. \ref{fig:orbit}, We select four sets of parameters and compare the orbital evolution in the x-y plane for different mass-ratios in each group. We can see that the  difference between the EOB orbit  and  the test particle one  is obvious, especially when the mass-ratio takes $10^{-4} $ and $10^{-5} $.
Note that the four orbits  we choose here are all zoom-whirl   \cite{PhysRevLett.103.131101}.
 This characteristic behavior involves several revolutions around the central body near the periastron \cite{PhysRevD.66.044002}, which leads to more pronounced orbital deviations than in generic orbits.

%相位图
\begin{figure} 
    \centering
    \begin{subfigure}[$a=0.99,e=0.9,\iota=0^{\circ},p=2.35 M$]{
	\includegraphics[scale=0.6]{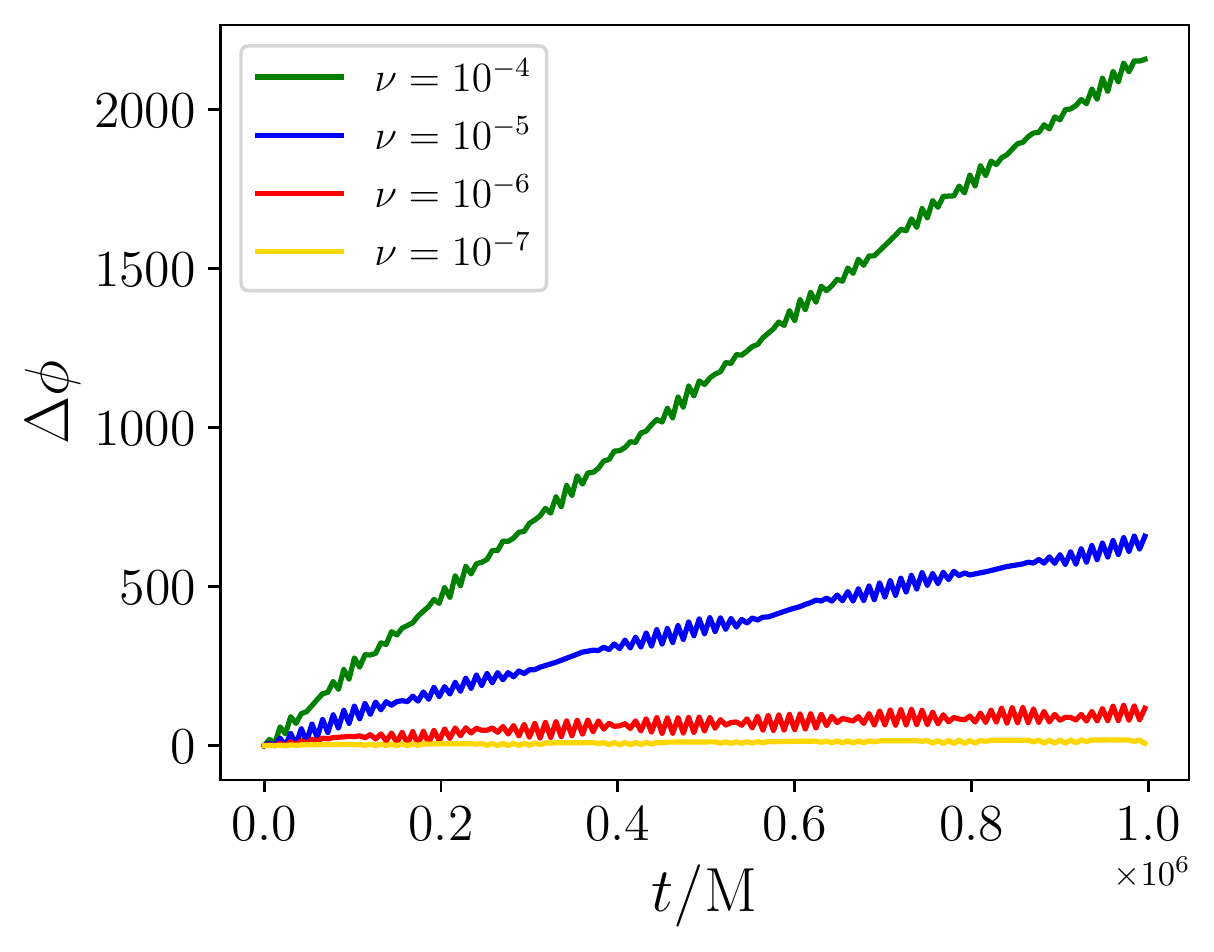}         
} 
	\end{subfigure}  
    \begin{subfigure}[$a=0.95,e=0.9,\iota=30^{\circ},p=3.15 M$]{
	\includegraphics[scale=0.6]{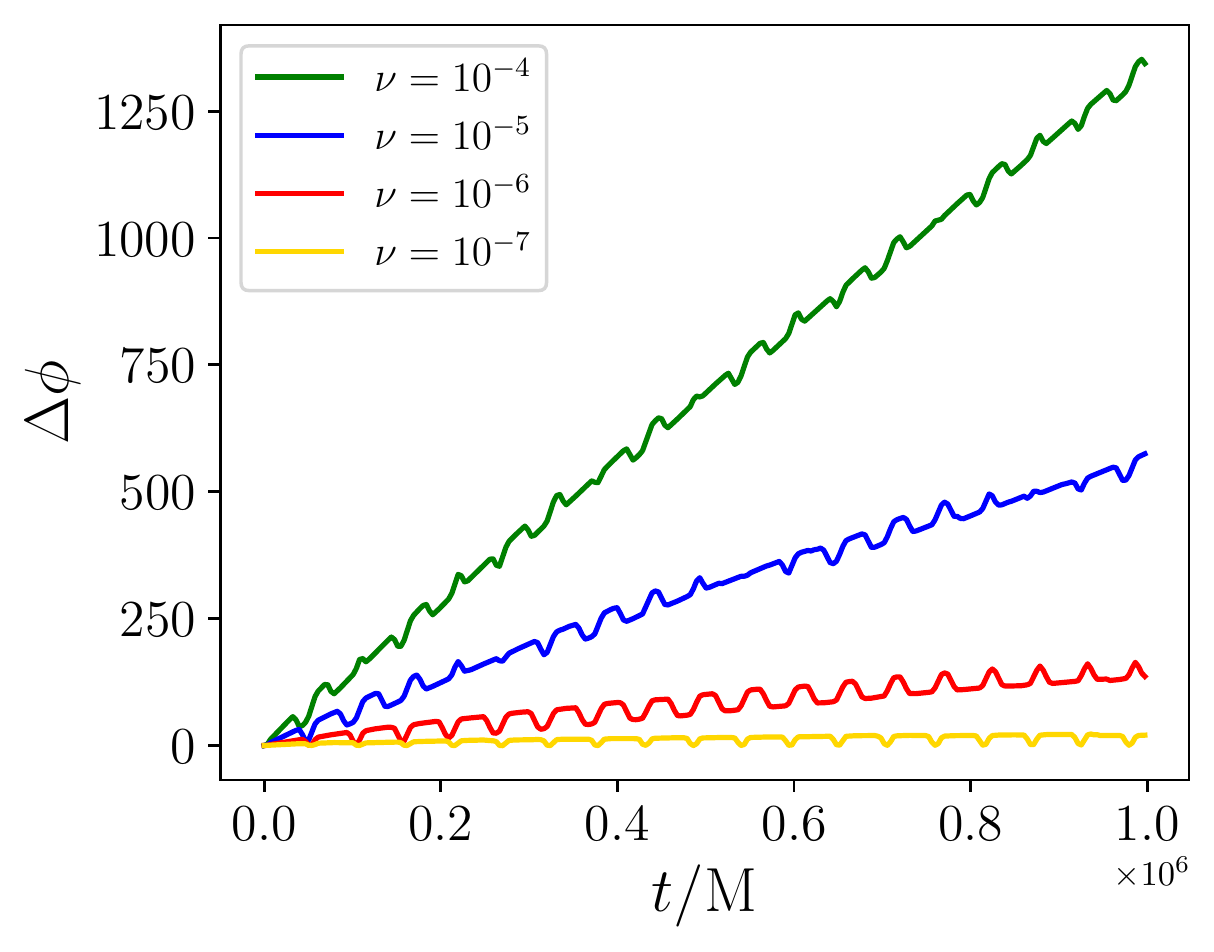}  
} 
        \end{subfigure}  
        
\begin{subfigure}[$a=0.9,e=0.9,\iota=10^{\circ},p=3.39 M$ ]{
	\includegraphics[scale=0.6]{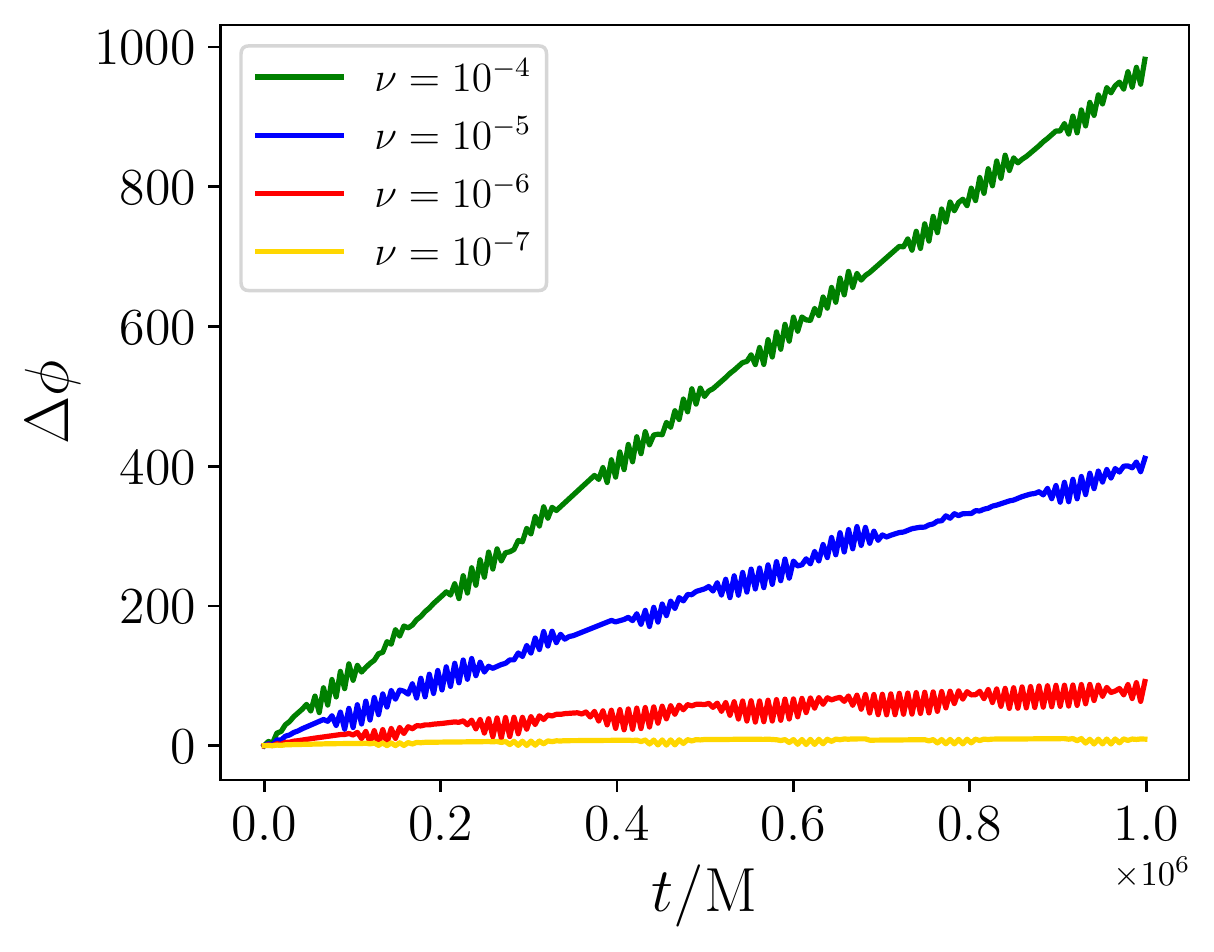} 
} 
	\end{subfigure}  	
 \begin{subfigure}[$a=0.98,e=0.8,\iota=0^{\circ},p=2.41 M$]{
	\includegraphics[scale=0.6]{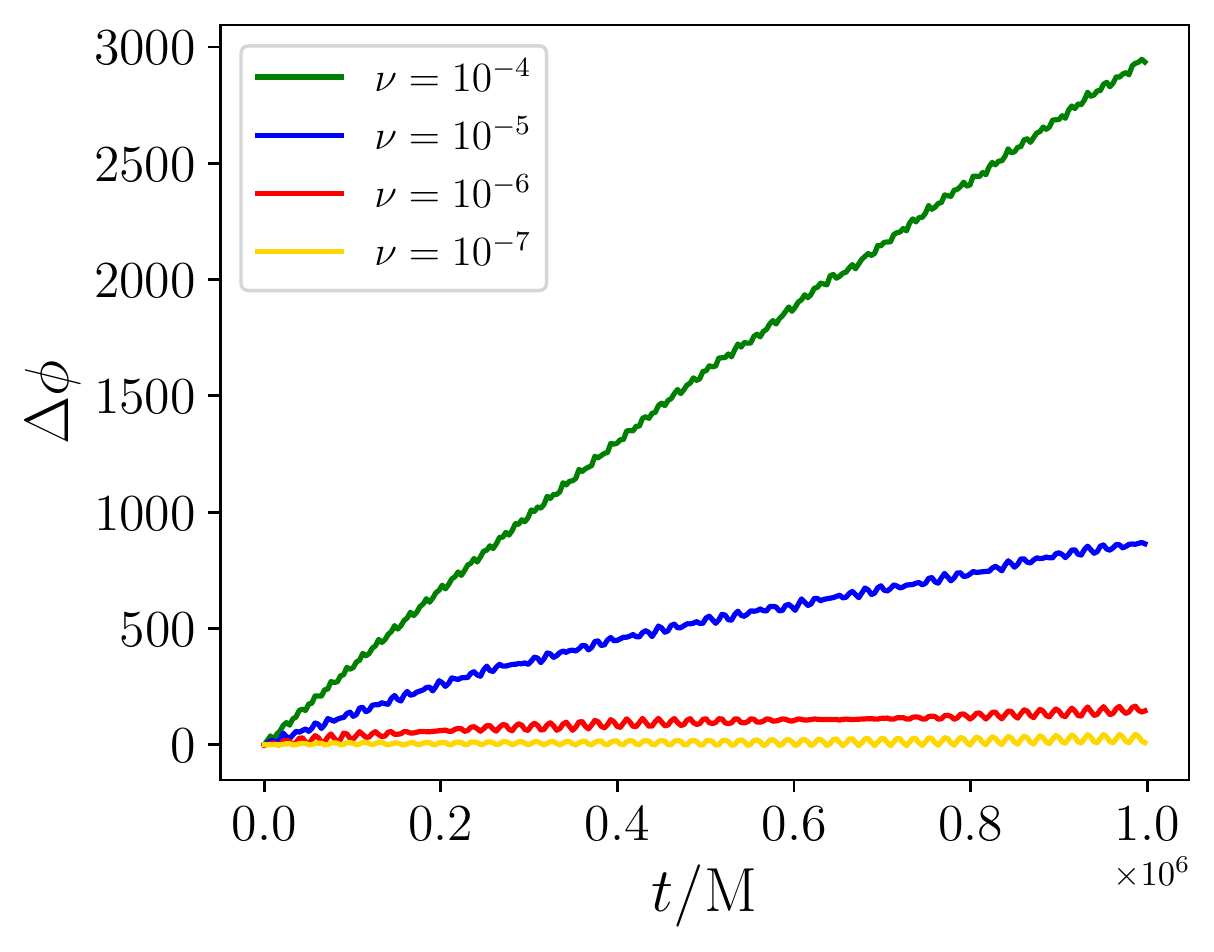}  
} 
	\end{subfigure}  
    \caption{\label{fig:delta phi}  Dephasing  ($\Delta \phi = \phi_\nu - \phi_{\nu = 0} $)  for different orbital parameters. The green, blue, red, and yellow  lines correspond to the cases: $\nu=10^{-4},10^{-5},10^{-6} $, and $\nu=10^{-7} $, respectively.} 
\end{figure}  

Fig. \ref{fig:delta phi}  represents the dephasing, $\Delta \phi$, for the four orbits presented in Fig. \ref{fig:orbit}. For high values of $\nu$, the phase shift $\Delta \phi$ becomes significant.  It is crucial to understand that even though the dephasing may be minor at very low values of $\nu$  ($\nu = 10^{-7} $), it cannot be dismissed as negligible. This is because we require sub-radian dephasing throughout the entire inspiral for an accurate extraction of the signal. It is essential to consider even the smallest deviations in order to obtain reliable results. Furthermore, we observe that as time increases, the dephasing $\Delta \phi$ derived for different mass-ratios becomes progressively larger.

The  expressions of the coordinate-time frequencies $\omega_{r} $, $\omega_{\theta} $, and $\omega_{\phi} $, considering the mass-ratio corrections, are given by   \cite{zhang2021geometrized} 
\begin{subequations} \label{eq:eob-omega} 
    \begin{gather} 
        \omega_r=\frac{\pi K (k) } {K (k)  W +a^2 z^2_+ E[K (k) -E (k) ] X}  \label{eq:eob-omega-r},\\
        \omega_{\theta} =\frac{\pi \beta z_+ X} {2\{K (k)  W +a^2 z^2_+ E[K (k) -E (k) ] X\} }  \label{eq:eob-omega-theta}  ,\\
        \omega_{\phi} =\frac{K (k) Z +L_z[\Pi (z^2_- ,k) -K (k) ]X} {K (k)  W +a^2 z^2_+ E[K (k) -E (k) ] X}  .\label{eq:eob-omega-phi} 
    \end{gather} 
\end{subequations} 
The expressions of $K (k) $, $E (k) $, $\Pi (z^2_- ,k) $, $X$, $z_+$, and $k$ are given in Appendix A (Eqs.  \eqref{eq:K}--\eqref{eq:k}) .

\begin{figure} 
    \centering
    \begin{subfigure}[$a=0.99,e=0.9,\iota=0^{\circ},p=2.35 M$]{
	\includegraphics[scale=0.6]{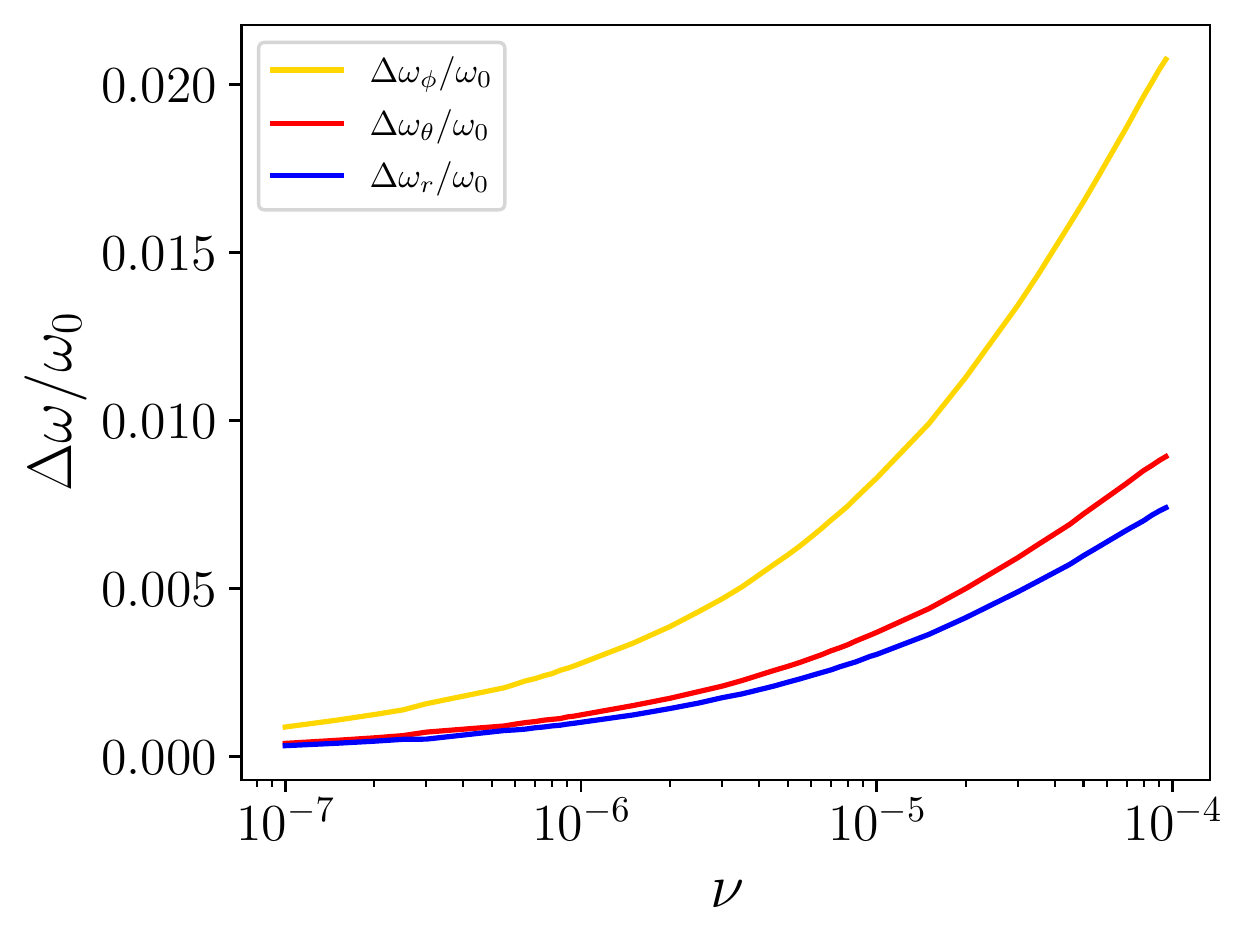}         
} 
	\end{subfigure}  
    \begin{subfigure}[$a=0.95,e=0.9,\iota=30^{\circ},p=3.15 M$]{
	\includegraphics[scale=0.6]{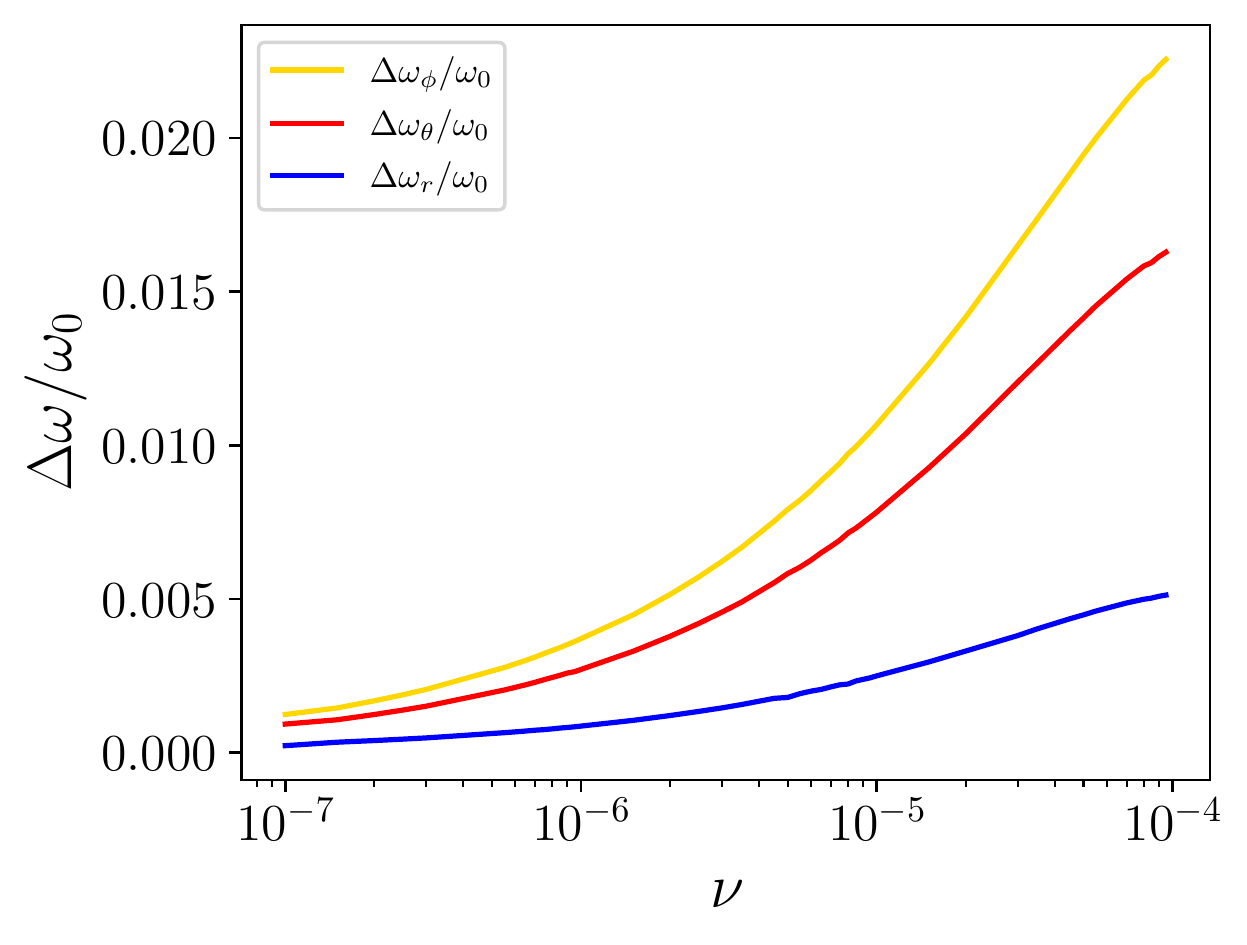}  
} 
        \end{subfigure}  
        
\begin{subfigure}[$a=0.9,e=0.9,\iota=10^{\circ},p=3.39 M$ ]{
	\includegraphics[scale=0.6]{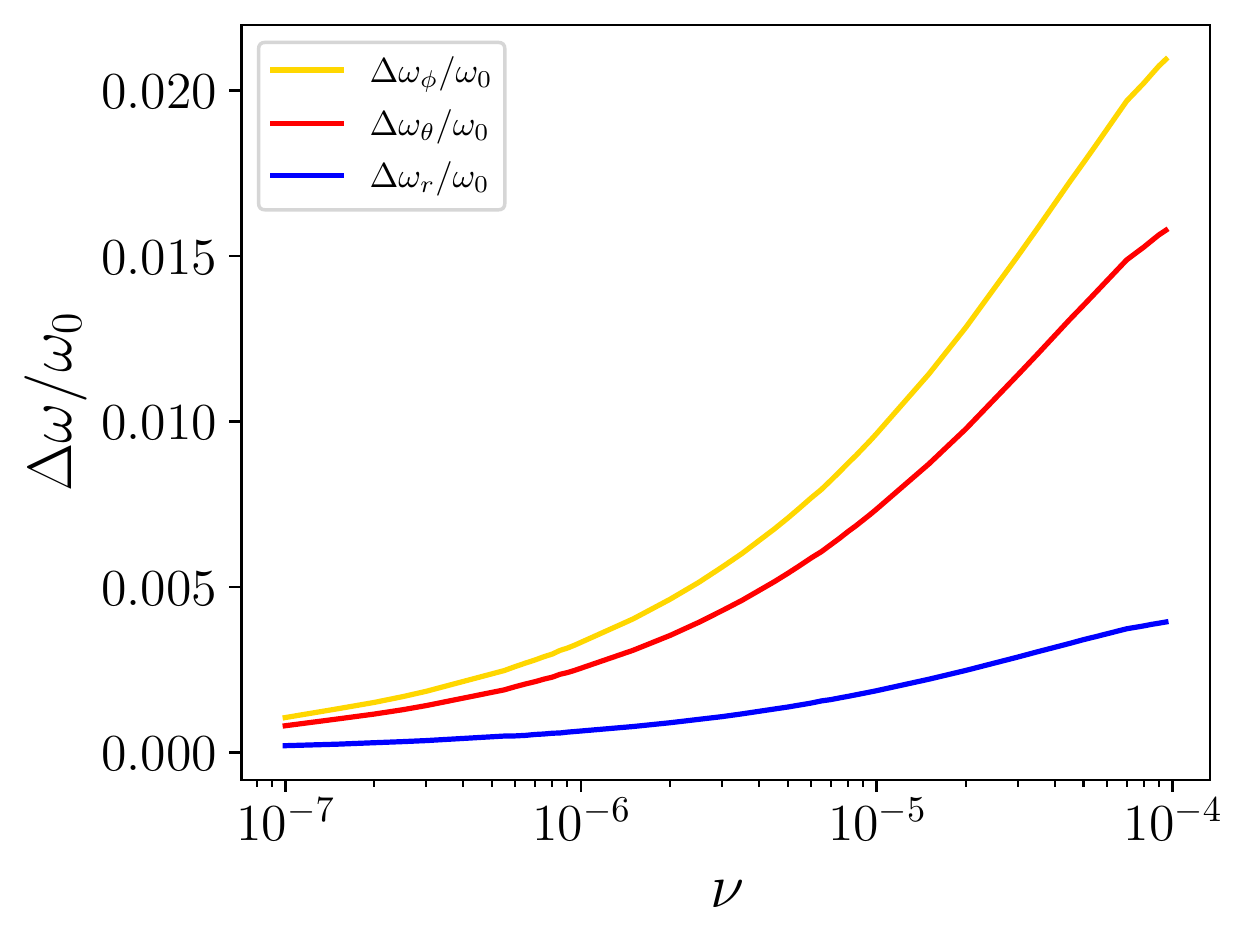} 
} 
	\end{subfigure}  	
 \begin{subfigure}[$a=0.98,e=0.8,\iota=0^{\circ},p=2.41 M$]{
	\includegraphics[scale=0.6]{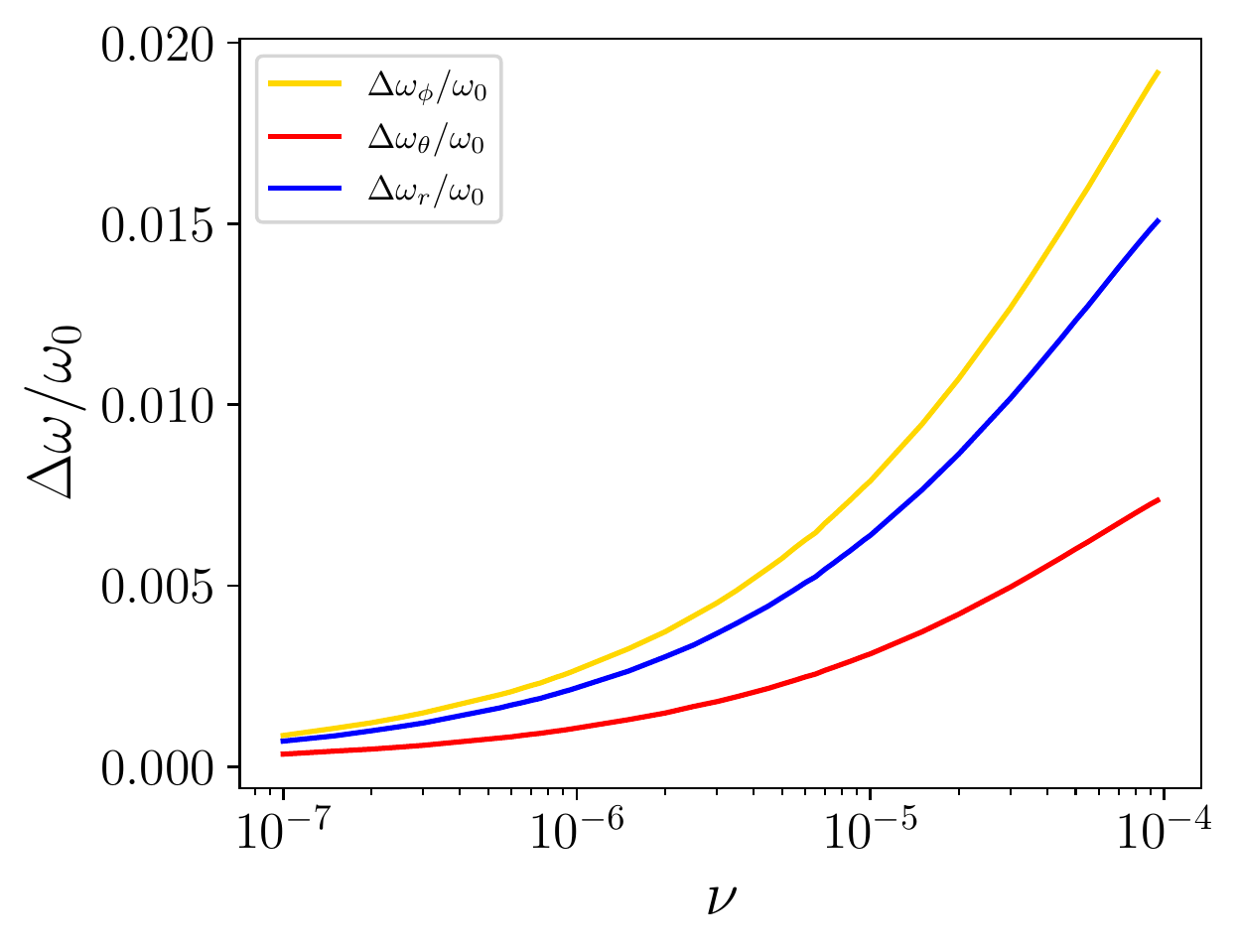}  
} 
	\end{subfigure}  
    \caption{\label{fig:frequency shift}  Frequency shifts ${\Delta {\omega} } /{ \omega_0} $  versus mass-ratio $\nu$ of four orbits with different orbital parameters. The yellow, red and blue lines represent $\Delta \omega_{\phi}  /  \omega_0 $, $\Delta \omega_ {\theta}  /  \omega_0 $ and  $\Delta \omega_{r}  /  \omega_0 $, respectively .} 
\end{figure} 
In Fig. \ref{fig:frequency shift}, we show the relative frequency shifts  $\Delta \omega/  \omega_0 = \mid\omega_\nu-\omega_0 \mid /\omega_0 $  (where the subscript 0 denotes the case of the test particle $ \nu=0 $)  of four  orbits with the same parameters in  Fig.  \ref{fig:orbit}. We find that the relative frequency shifts increase with the mass-ratio.  In addition, the relative frequency shifts can be up to about two orders  ($10^2\nu$)  larger than the mass-ratio for extremely relativistic orbits.

All the above results show that even in the case of extreme mass-ratio, the mass-ratio corrections will have a significant impact on orbital frequency, and then affect orbital evolution. This means that when constructing EMRIs waveform template, the mass-ratio should not be ignored. In other words, if we neglect the mass-ratio correction when building EMRIs waveform template, the final estimated parameters may induce a fake signal of deviation from GR, which  will be  discussed in more detail in Sec.\ref{bayes}. 

We point out that all the results presented in this section only take into account the conservative dynamics, in which the mass-ratio corrections are associated with the conservative part of the 1SF. 
Though the mass-ratio correction maybe not correspond to the exact conservative GSF, we can always improve the accuracy of the EOB model by calibrating it with  GSF results  \cite{Barack_2018,PhysRevD.106.064049,PhysRevD.106.084062}.
It is important to note that accurately modeling EMRI systems throughout the whole inspiral  requires 1PA corrections.
Besides the conservative part of 1SF,  the dissipative part up to  2SF is also essential and should be taken into account  \cite{PhysRevD.78.064028, PhysRevD.105.084031}.

\section{\label{waveforms}  waveforms and  orbital evolution} 

The gravitational waveforms can be calculated in the Teukolsky formalism, considering  the perturbation on the Weyl scalar $\psi_4$   \cite{1972PhRvL..29.1114T,teukolsky1973perturbations}.
At infinity, the GW polarizations $h_{+,\times} $ can be written as 
\begin{equation} 
    \psi_4 (r\rightarrow \infty) =\frac{1} {2} \frac{\partial^2} {\partial t^2}  (h_+-ih_{\times} ).
\end{equation} 
We decompose $\psi_4$ in the frequency domain
\begin{equation} 
    \psi_4=\rho^4 \int_{-\infty} ^{+\infty}  d \omega \sum _{lm}  R_{lm\omega}  (r)  _{-2} S_{lm} ^{a\omega}  (\Theta)  e^{im\Phi}  e^{-i\phi},
\end{equation} 
where $\rho=-1/ (r-i a \cos{\theta} ) $, $\omega$ is the  discrete frequency spectrum, and $\phi=\int \omega dt $.  Here, $_{-2} S_{lm} ^{a\omega }  (\Theta) $ denotes the spin-weighted ($s=-2$) spherical harmonic coefficient, which depends on the polar angle $\Theta$.
The function $R_{lm\omega}  (r) $ satisfies the radial Teukolsky equation
\begin{equation} \label{eq:Rlmkn} 
    \Delta^2 \frac{d} {dr}  (\frac{1} {\Delta}  \frac{d R_{lm\omega} } {dr} ) -V (r) R_{lm\omega} =-\mathscr{T} _{lm\omega}  (r),
\end{equation} 
where $\Delta= r^2 -2 Mr + a^2$,  $V (r)$ is the potential (Eq. (2.5) of Ref. \cite{10.1143/PTPS.128.1}), and $\mathscr{T}$ is the source term  (Eq. (2.14) of Ref. \cite{10.1143/PTPS.128.1}).
The general solution of Eq. \eqref{eq:Rlmkn}  is
\begin{equation} 
    R_{lm\omega}  (r) =\frac{R^{\infty} _{lm\omega}  (r) } {2i\omega B^{\rm in} _{lm\omega} D^{\infty} _{lm\omega} } \int_{r_+} ^r dr' \frac{R^{\rm H} _{lm\omega}  (r')  \mathscr{T} _{lm\omega}  (r') } {\Delta (r') ^2} 
    +\frac{R^{\rm H} _{lm\omega}  (r) } {2i\omega B^{\rm in} _{lm\omega} D^{\infty} _{lm\omega} } \int_{r} ^{\infty}  dr' \frac{R^{\infty} _{lm\omega}  (r')  \mathscr{T} _{lm\omega}  (r') } {\Delta (r') ^2},
\end{equation} 
%\red{where $B^{\rm in} _{lm\omega}$ is the asymptotic amplitudes, and D is ?.}
where $B^{\rm in} _{lm\omega}$ and $D^{\infty} _{lm\omega}$ are the asymptotic amplitudes \cite{Fujita_2005,PhysRevD.102.024041}. %, and it is convenient to fix their values as shown in Appendix A of  \cite{PhysRevD.102.024041}.
$R^{\infty} _{lm\omega}  (r) $ and $R^{\rm H} _{lm\omega}  (r) $ are the two independent solutions of the homogeneous Teukolsky equation, which are given by
\begin{equation} 
    R^{\infty} _{lm\omega}  (r) =Z^{\rm H} _{lm\omega}  r^3 e^{i\omega r^*},
\end{equation} 
\begin{equation} 
    R^{\rm H} _{lm\omega}  (r) =Z^{\infty} _{lm\omega}  \Delta^2 e^{-i p r^*},
\end{equation} 
with 
\begin{equation} 
    Z^{\infty} _{lm\omega} =\frac{B^{\rm hole} _{lm\omega} } {2i\omega B^{\rm in} _{lm\omega} D^{\infty} _{lm\omega} } \int_{r} ^{\infty}  dr' \frac{R^{\infty} _{lm\omega}  (r')  \mathscr{T} _{lm\omega}  (r') } {\Delta (r') ^2},
\end{equation} 
\begin{equation} 
    Z^{\rm H} _{lm\omega} =\frac{1} {2i\omega B^{\rm in} _{lm\omega} } \int_{r_+} ^r dr' \frac{R^{\rm H} _{lm\omega}  (r')  \mathscr{T} _{lm\omega}  (r') } {\Delta (r') ^2},
\end{equation} 
where $r^*$ is the tortoise coordinate. The amplitudes $Z^{\rm {H},\infty} _{lm\omega}  (r) $ fully determine the energy and angular momentum fluxes $\Dot{E} ^{{\rm H},\infty} $ , $\Dot{L} _z^{{\rm H},\infty} $ and the gravitational waveforms.
\begin{equation} 
    \Dot{E} ^{\infty,{\rm H} } =\sum_{lm \omega}  \frac{\mid Z^{{\rm H},\infty} _{lm\omega}  \mid ^2} {4\pi \omega^2},
\end{equation} 

\begin{equation} 
    \Dot{L} _z^{\infty,{\rm H} } =\sum_{lm \omega}  \frac{m \mid Z^{{\rm H},\infty} _{lm\omega} \mid^2} {4\pi \omega^3}.
\end{equation}

\begin{figure} 
    \centering
	\includegraphics[scale=0.7]{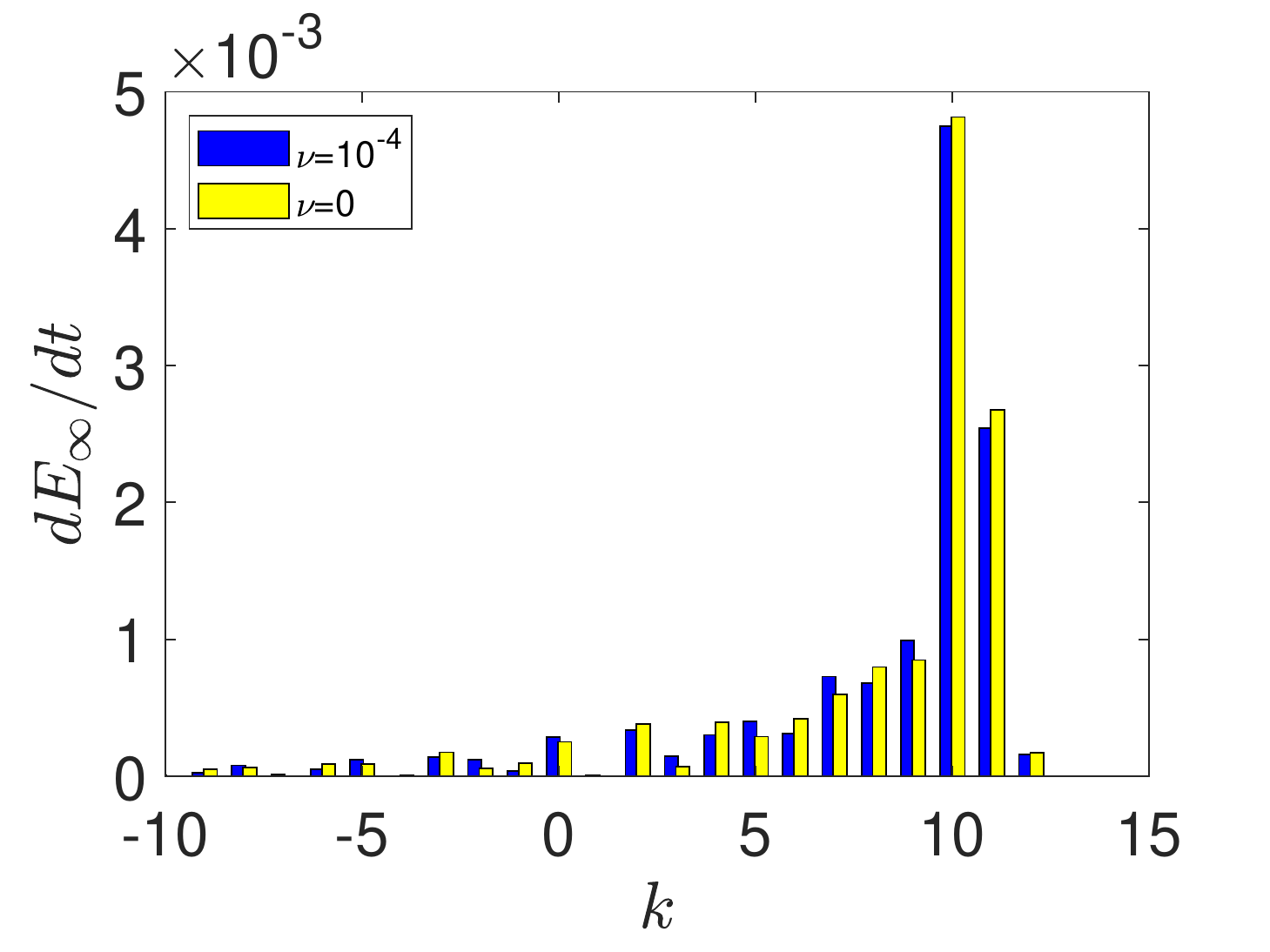}         
    \caption{\label{fig:dedtnu}  The spectrum of energy fluxes versus the harmonic number $k$ with parameters $a = 0.99$, $p = 2.11$ and $e = 0.7$. The yellow and blue bar represent $\nu = 0$ and $\nu  =10^{-4} $, respectively.
    }
    %\red{We note that the yellow bar is not always above the blue one?.} 
\end{figure}

Due to the modification of the orbit by the mass-ratio, the orbital averaged energy flux between the EOB orbit and the test particle one could be different obviously. Fig.  \ref{fig:dedtnu}  shows that for a
mass ratio of $10^{-4} $, the energy flux spectrum is visibly different between both
cases. Our calculations from the Teukolsky equation show that for the relativistic orbits, the relative difference of the energy fluxes between the EOB and test particle orbits is up to 10$\nu$ magnitude. This will produce enough GW dephasing during the long evolution of EMRIs.   

For the generic orbits, the particle's motion can be described as the harmonic of the frequency $\omega_{\phi} $,  $\omega_{\theta} $, and $\omega_{r} $. We take $\omega=\omega_{mkn} = m \omega_{\phi} + k \omega_{\theta}  + n \omega_{r} $, $\phi_{mkn} =\int \omega_{mkn}  dt $.
Therefore, we can write $h_{+,\times} $ as the multipolar sum of ``voices" with these frequencies   \cite{1974ApJ...193..443T}.
For an EMRI system with a total mass $M$ at a distance $D$, an altitude angle $\Theta$ and an azimuth angle $\Phi$,
its GW strain can be written as    \cite{Pound_2021,Hughes_2021} 
\begin{equation} \label{eq:strain} 
    h_+-ih_{\times} =-\frac{2\mu} {D} \sum_{lmkn}  \frac{Z_{lmkn} ^{\infty} } {\omega^2_{mkn} } 
    \frac{_{-2} S_{lm} ^{a\omega_{mkn} }  (\Theta) } {\sqrt{2\pi} } 
    e^{-i\phi_{mkn} +im\Phi},
\end{equation} 
where $\sum_{lmkn} =\sum_{l=2} ^{\infty}  \sum_{m=-l} ^{l}   \sum_{k=-\infty} ^{\infty}  \sum_{n=-\infty} ^{\infty} $, and the amplitude $Z_{lmkn} ^{\infty} $ of each mode  can be calculated using the radial Teukolsky equation. Here, $_{-2} S_{lm} ^{a\omega_{mkn} }  (\Theta) $ denotes the spin-weighted ($s=-2$)  spherical harmonic coefficient, which depends on the polar angle $\Theta$.

In the present work, we combine the EOB formalism with the Teukolsky equation, i.e. the EOB trajectory sources the Teukolsky equation and the latter calculates the waveforms and feeds back the EOB orbit. The gravitational waveforms are generated by using Eq. \eqref{eq:strain}, which is the multipolar sum of ``voices" with orbital frequencies $\omega_{\phi} $,  $\omega_{\theta} $, and $\omega_{r} $. Since the time scale of gravitational radiation of EMRIs is far larger than the orbital period, by using  the adiabatic approximation,  we can calculate these frequencies with Eqs. \eqref{eq:eob-omega-r}--\eqref{eq:eob-omega-phi}.
Unlike the case of the test particle, the EOB dynamics includes the first-order conservative GSF corrections. Note that it maybe not entirely exact but we can always improve the accuracy of the EOB model by calibrating it with  GSF results \blue{\cite{Barack_2018,PhysRevD.106.064049,PhysRevD.106.084062}}.

Considering the gravitational radiation, the energy, angular momentum, and the Carter constant are no longer conserved. The rates of their changes are expressed by

\begin{subequations} \label{eq:dot} 
\begin{align} 
\frac{\mathrm{d}  E} {\mathrm{d}  t} =\frac{\partial E} {\partial p} \Dot{p} +\frac{\partial E} {\partial e} \Dot{e} +\frac{\partial E} {\partial \iota} \Dot{\iota} \label{eq:Edot},\\
\frac{\mathrm{d}  \hat{L} _z} {\mathrm{d}  t} =\frac{\partial \hat{L} _z} {\partial p} \Dot{p} +\frac{\partial  \hat{L} _z} {\partial e} \Dot{e} +\frac{\partial  \hat{L} _z} {\partial \iota} \Dot{\iota} \label{eq:Lzdot},\\
\frac{\mathrm{d}  \hat{Q} } {\mathrm{d}  t} =\frac{\partial \hat{Q} } {\partial p} \Dot{p} +\frac{\partial \hat{Q} } {\partial e} \Dot{e} +\frac{\partial \hat{Q} } {\partial \iota} \Dot{\iota} \label{eq:Qdot1}.
\end{align} 
\end{subequations} 
 By solving Eqs.  \eqref{eq:Edot} -\eqref{eq:Qdot1}, we can describe the evolution  of orbital parameters $ (p,e,\iota) $ by
\begin{subequations} \label{eq:p,e,iota} 
\begin{align} 
\Dot{p} =c_{ (\hat{L} _z,\hat{Q} )  (e,\iota) } \frac{\mathrm{d}  E} {\mathrm{d}  t} +c_{ (E,\hat{Q} )  (\iota,e) } \frac{\mathrm{d}  \hat{L} _z} {\mathrm{d}  t} +c_{ (E,\hat{L} _z)  (e,\iota) } \frac{\mathrm{d}  \hat{Q} } {\mathrm{d}  t} \label{eq:p-dot},\\
\Dot{e} =c_{ (\hat{L} _z,\hat{Q} )  (\iota,p) } \frac{\mathrm{d}  E} {\mathrm{d}  t} +c_{ (E,\hat{Q} )  (p,\iota) } \frac{\mathrm{d}  \hat{L} _z} {\mathrm{d}  t} +c_{ (E,\hat{L} _z)  (\iota,p) } \frac{\mathrm{d}  \hat{Q} } {\mathrm{d}  t} \label{eq:e-dot},\\
\Dot{\iota} =c_{ (\hat{L} _z,\hat{Q} )  (p,e) } \frac{\mathrm{d}  E} {\mathrm{d}  t} +c_{ (E,\hat{Q} )  (e,p) } \frac{\mathrm{d}  \hat{L} _z} {\mathrm{d}  t} +c_{ (E,\hat{L} _z)  (p,e) } \frac{\mathrm{d}  \hat{Q} } {\mathrm{d}  t} \label{eq:iota-dot},
\end{align} 
\end{subequations} 
where the coefficients are given by 
\begin{equation} 
    c_{ (C_1,C_2)  (o_1,o_2) } =\frac{\frac{\partial C_1} {\partial o_1}  \frac{\partial C_2} {\partial o_2} -\frac{\partial C_1} {\partial o_2} \frac{\partial C_2} {\partial o_1}   } 
    { 
    [\frac{\partial E} {\partial \iota}  \frac{\partial \hat{L} _z} {\partial p} -\frac{\partial E} {\partial p} \frac{\partial \hat{L} _z} {\partial \iota} ]
    \frac{\partial \hat{Q} } {\partial e} 
    +
    [\frac{\partial E} {\partial e}  \frac{\partial \hat{L} _z} {\partial \iota} -\frac{\partial E} {\partial \iota} \frac{\partial \hat{L} _z} {\partial e} ]
    \frac{\partial \hat{Q} } {\partial p} 
    +
    [\frac{\partial E} {\partial p}  \frac{\partial \hat{L} _z} {\partial e} -\frac{\partial E} {\partial e} \frac{\partial \hat{L} _z} {\partial p} ]
    \frac{\partial \hat{Q} } {\partial \iota} 
    }.
\end{equation} 

Once we have the GW  fluxes $\Dot{E} $, $\Dot{L} _z$, and $\Dot{Q} $, substituting  expressions of these  fluxes  into Eqs.  \eqref{eq:p-dot}--\eqref{eq:iota-dot}, we can derive the orbital evolution.
The $_{-2} S_{lm} ^{a\omega_{mkn} }$ and $Z_{lmkn} ^{\infty}$ can be computed by using the analytical module of the Black Hole Perturbation Club (BHPC) code  \cite{Black-Hole-Perturbation-Club,PhysRevLett.128.231101,10.1143/PTP.121.843}.
Combing Eqs.  \eqref{eq:eob-omega-r}--\eqref{eq:eob-omega-phi}  and Eq. \eqref{eq:strain}, we can generate the gravitational waveforms.
However, fully using the numerical Teukolsky-based fluxes is computationally expensive. 
Without impacting  our main target, we here employ a hybrid scheme of fluxes  \cite{gair2006improved}.

Compared  to the numerical fluxes, the GG formalism performs well for orbits of low  eccentricity, especially  for circular, inclined orbits.  We also include mass-ratio corrections (dissipative 1PA term) to the original 2PN fluxes based on  \cite{PhysRevD.96.104048,Zhang_2021,10.1143/PTP.121.843}, though this is almost no influence on the data analysis results. The final expressions for $\Dot{E} $, $\Dot{L} _z$, and $\Dot{Q} $ (Eqs. \eqref{eq:Lz mod}--\eqref{eq:E mod} with Eqs. \eqref{eq:Lzfit}--\eqref{eq:iotafit} and Eqs. \eqref{eq:newEdot2pn}--\eqref{eq:newQdot} )   are given in Appendix B. 
\begin{figure} 
    \centering
	\includegraphics[scale=0.6]{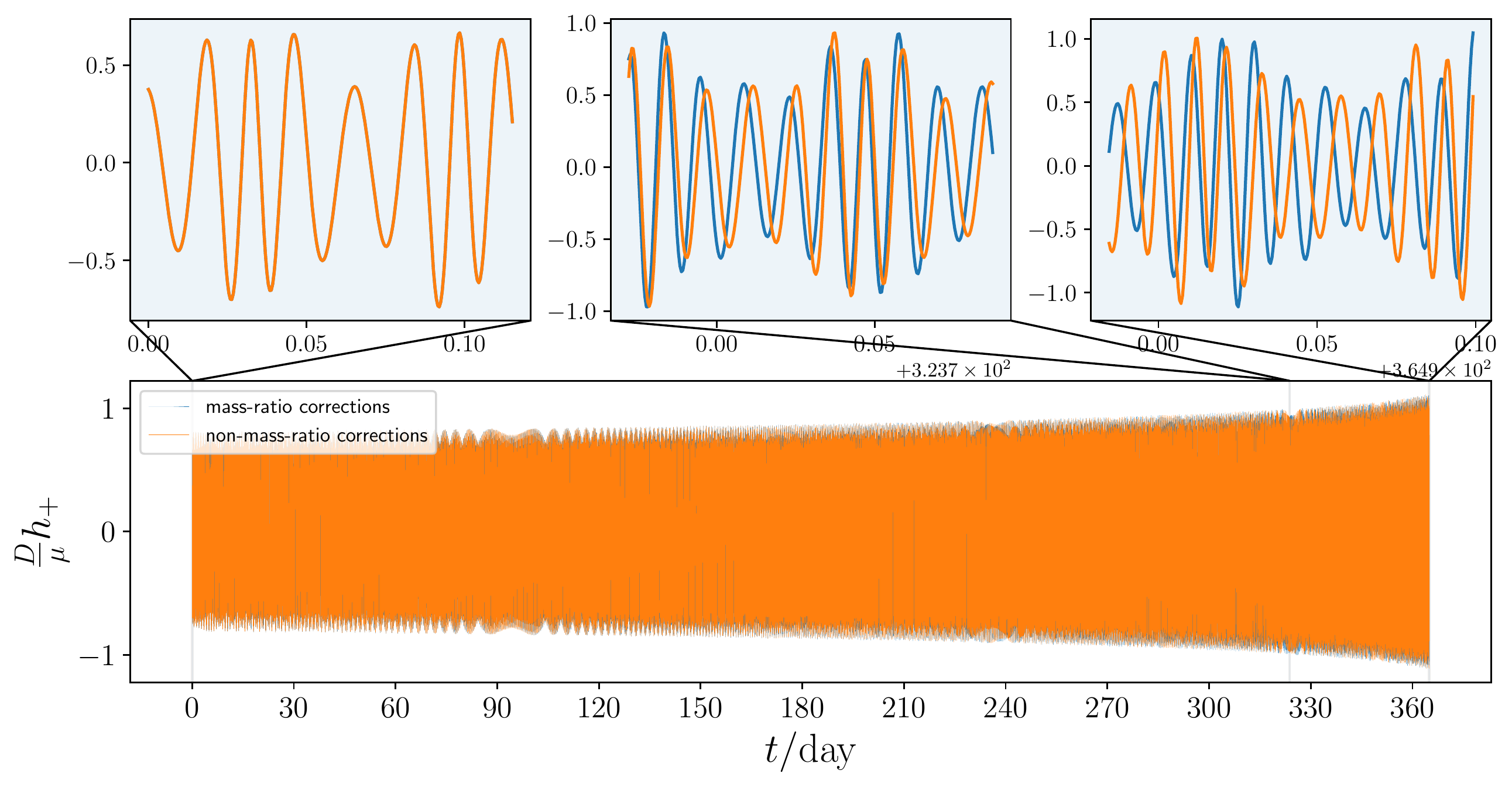}         
    \caption{\label{fig:strain}  Two waveforms of  EMRI systems with     
    $ (M, \mu, a )  = (5\times10^6 M_{\odot}, 50 M_{\odot},0.43) $, starting at $ (p_0, e_0, \iota_0)  = (6.84, 0.166, 45^\circ) $.  We  plot the one-year  evolving waveform at the viewing angle $ (\Theta, \Phi)  = (45^\circ, 0^\circ) $. The orange line represents the waveform excluding mass-ratio corrections, and the blue line denotes the waveform including mass-ratio corrections.   The waveforms are generated using the Black Hole Perturbation Club’s Teukolsky code  \cite{PhysRevLett.128.231101,Black-Hole-Perturbation-Club}. } 
\end{figure} 

As an example, Fig. \ref{fig:strain}  shows two waveforms  including or excluding mass-ratio corrections  starting at $ (p_0, e_0, \iota_0)  = (6.84, 0.166, 45^\circ) $, and the mass-ratio is $10^{-5} $. We plot the one-year evolving waveforms at the viewing angle $ (\Theta, \Phi)  = (45^\circ, 0^\circ) $.  In our work, we generate the gravitational waveforms only considering $ (l,m) = (2,2) $ mode, which is the dominant strain mode. 
The bottom row and the top one represent the one-year evolving waveforms and their three zoom-in waveforms, respectively.
From magnified waveforms (top row), we can see that at the beginning (left column), the difference between two gravitational waveforms with or without mass-ratio corrections is quite small, which is invisible to the naked eyes. 
However, waveform errors due to the mass-ratio can accumulate as time goes by. And then  obvious waveform differences (both in dephasing and amplitude) can be found in the waveforms at the latter stage of evolution (middle and right columns).

The accumulative errors are caused by the dual effects of mass-ratio correction on the orbital frequencies and the Teukolsky-based energy fluxes. First of all, under the same orbital parameters, mass-ratio will affect the orbital frequencies and thus impact the frequencies of GWs at $O (\nu) -O (10^2 \nu) $ depending on the orbital parameters. Furthermore, the orbit correction caused by mass-ratio will change the Teukolsky-based energy fluxes with a relative difference above $O (\nu) $ for relativistic trajectory. Under the joint influence of these two factors, when the mass-ratio correction is not considered in the orbit dynamics, i.e., calculate the geodesic in the background field of the massive body and then evolve the orbit by the adiabatic approximation  (the radiation reaction is considered as the leading order self-force), it may produce accumulative errors that can not be ignored for EMRIs waveforms.

\section{\label{bayes} data analysis and results} 

The matched filtering   \cite{PhysRevD.46.5236}  technique is widely used in LIGO and Virgo data processing, and it will also be applicable to future space-borne GW detectors. We also apply this technique to analyze the influence of mass-ratio corrections on EMRIs waveforms quantitatively.
Given two time series $a (t) $ and $b (t) $, their maximized fitting factor (overlap)  and mismatch ($\mathcal{M} $) are given by
\begin{equation} 
    {\rm{overlap} } = \max_{t_s,\phi_s} \frac{ (a (t) |b (t+t_s) e^{i\phi_s} ) } {\sqrt{ (a|a)  (b|b) } },
\end{equation} 
\begin{equation} \label{eq:mismatch} 
    {\mathcal{M} } =1-{\rm{overlap} } =1-\max_{t_s,\phi_s} \frac{ (a (t) |b (t+t_s) e^{i\phi_s} ) } {\sqrt{ (a|a)  (b|b) } },
\end{equation} 
where $t_s$ is the time shift, $\phi_s$ is the phase shift, 
and $ (a|b) $ is the inner product between two time series signal $a (t) $ and $b (t) $ with the following expression
\begin{equation} 
    (a|b) =2\int_0^\infty df \frac{\Tilde{a} ^* (f)  \Tilde{b}  (f) +\Tilde{a}  (f)  \Tilde{b} ^* (f) } {S_n (f) }  ,
\end{equation} 
where $\Tilde{a}  (f) $ is the Fourier transform of the series signal $a (t) $, $\Tilde{a} ^* (f) $ is the complex conjugate of $\Tilde{a}  (f) $,
and $S_n (f) $ is the power spectral density (PSD). In this work, we use the PSD from the  Online Sensitivity Curve Generator website    \cite{Online-Sensitivity-Curve-Generator,PhysRevD.66.122002}, representative of the design sensitivity of LISA.

\begin{figure} 
    \centering
	\includegraphics[scale=0.7]{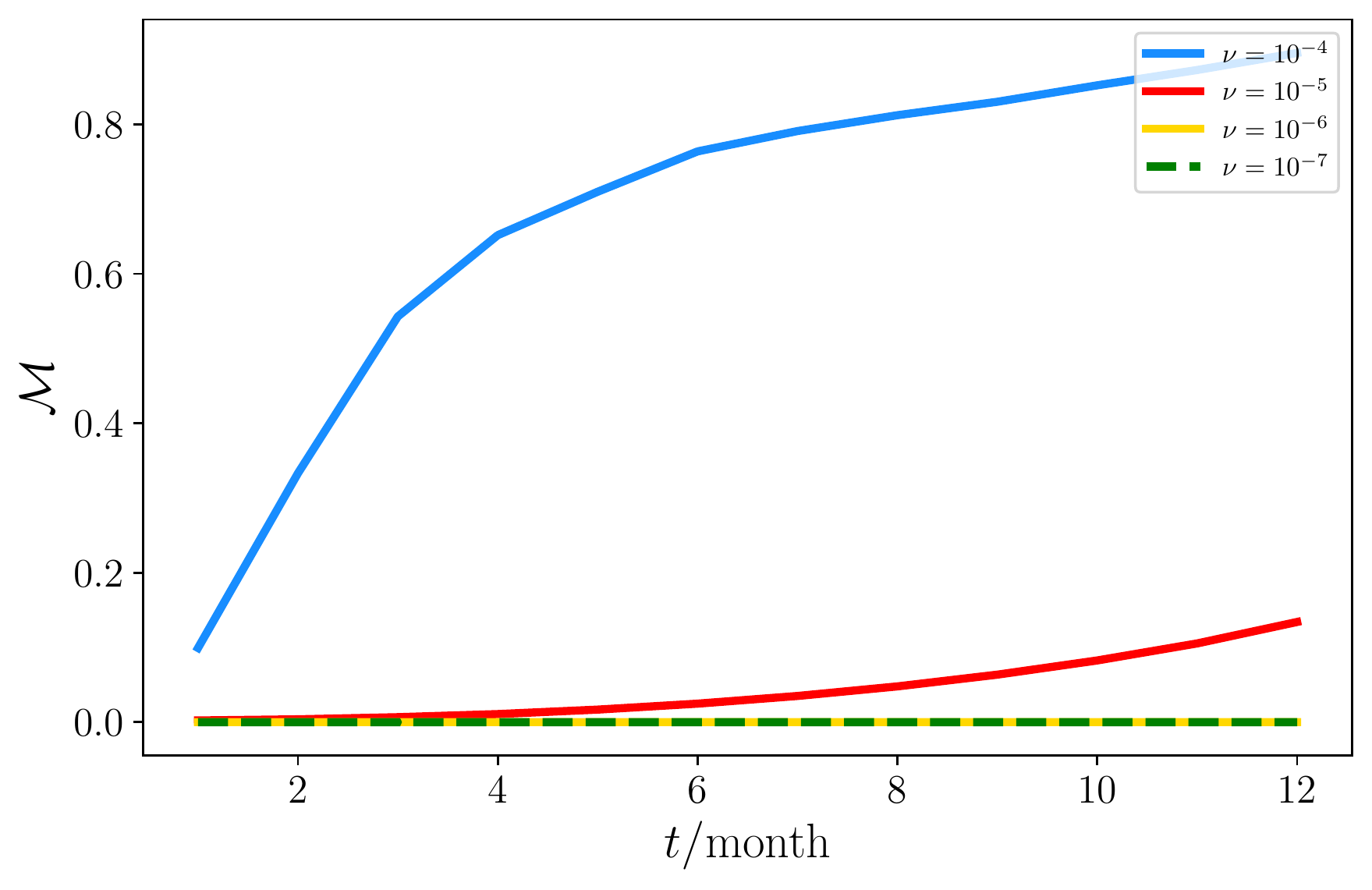}         
    \caption{\label{fig:mismatch} The mismatches of waveforms coming from EOB dynamical evolutions and the test particle one.
    The blue, red, yellow, and green lines  represent systems with different mass-ratios of $10^{-4} $, $10^{-5} $, $10^{-6} $, and  $10^{-7} $,  respectively.
    For each $\nu$, we calculate the mismatch considering different evolution times from 1 month to 12 months. We fix the EMRI parameters to $a=0.9$, $ (\mu, M )  = ( 50 M_{\odot}, \mu/\nu ) $ at the distance of $1\rm{Gpc} $. Our waveforms end at $ ( p, e, \iota )  = (5.1, 0.1, 80^\circ) $. } 
\end{figure}

We use the PyCBC library    \cite{alex_nitz_2022_6912865}  to compute the mismatch between two EMRI waveforms at a distance of 1 Gpc, and the result is presented in Fig. \ref{fig:mismatch}. 
We fix the mass of the SCO at $50 M_{\odot} $ and consider four EMRI systems with mass-ratios ranging from $10^{-4} $ to $10^{-7} $. The results show that when the mass-ratio ($\nu=10^{-6} $ and $\nu=10^{-7} $) is closer to the test particle limit ($\nu=0$), the mismatch is so small that  can be ignored. In addition, for the case of $\nu=10^{-4} $ and $\nu=10^{-5} $,  as time passes, the mismatches between two waveforms will increase  and may be large enough to induce a fake signal  of deviation from GR, which will be explained  later.

We follow  the method of Moore   \cite{moore2021testing}  et al.  to calculate fake Bayesian factors for the deviation from GR. First, it is assumed that GR is the correct description of nature. For the combined parameter space $\lambda= (\alpha;\theta) $, 
the observed signal, s, can be described as the sum of the GW signal and the detector noise.
 \begin{equation} 
 \begin{split} 
     s&=n+h (\alpha;\theta)  \\ 
      &=n+h (\alpha_{\rm Tr} =0;\theta_{\rm Tr} ) +\Delta h (\theta_{\rm Tr} ).
 \end{split} 
 \end{equation} 
 Here, we treat $\alpha$ as the modified gravity parameter, and in the case of GR, $\alpha$ is equal to 0. $\theta$ denotes the  source parameters including both intrinsic (masses, spins, etc.) and extrinsic (distance, viewing angles, etc.) parameters. $\Delta h (\theta_{\rm Tr} ) $ represents the errors caused by inaccurate waveform templates.

Assuming the instrument noise is Gaussian, the likelihood  $\mathcal{L}  (\alpha;\theta) =P (s|\alpha;\theta) $ is 
\begin{equation} 
 \begin{split} 
     \log \mathcal{L}  (\alpha;\theta) &=-\frac{1} {2}  \mid s-h (\alpha;\theta) \mid ^2+c\\
     &=-\frac{1} {2}  \mid n-\delta h (\alpha;\theta) +\Delta h (\theta_{\rm Tr} ) \mid ^2+c,
 \end{split} 
 \end{equation} 
where we set $\delta h (\alpha;\theta) =h (\alpha;\theta) -h (\alpha_{\rm Tr} ;\theta_{\rm Tr} ) $  and $c$ is a constant. For simplicity, we assume that the prior on $\lambda$ is flat, so the posterior is proportional to the likelihood. 
The maximum likelihood (ML)  parameters can be decomposed into the following three parts
 \begin{equation} 
     \lambda_{\rm ML} =\lambda_{\rm Tr} + \Delta\lambda_{\rm stat}  +  \Delta\lambda_{\rm sys},
 \end{equation} 
 where  $\lambda_{\rm Tr}  $ is the true source parameters, $\Delta\lambda_{\rm stat} $ is the statistical error, which depends on the instrument noise $n$, and $\Delta\lambda_{\rm sys} $ is the systematic error, which  depends on  the model error $\Delta h$. Through the first-order Taylor expansion of $\Delta h$, the log-likelihood at the ML parameters is given by   \cite{moore2021testing}  
\begin{equation}  \label{eq:log likelihood} 
    \log \mathcal{L}  (\lambda) =c'-\frac{1} {2} \Gamma_{\mu \nu}  (\lambda-\lambda_{\rm ML} ) ^{\mu}  (\lambda-\lambda_{\rm ML} ) ^{\nu},
\end{equation} 
where  $c'$ is another constant, $\Gamma_{\mu \nu} $ is the Fisher matrix and we assume that  $\Gamma_{oi} =0$ if $i\neq 0$.
We set $k=\rm{dim}  (\theta) $, that is, ${\rm dim }  (\lambda) =k+1$.
For the non-GR sub-model, The Bayesian evidence integral is given by
\begin{equation} 
    Z_{\rm non-GR} =\int d\lambda \mathcal{L} _ (\lambda) =e^{c'} 
    \sqrt{\frac{ (2\pi) ^{k+1} } {\det  \Gamma_{\mu \nu} } }.
\end{equation} 
For the GR sub-model, $\alpha=0$, we have 
\begin{equation} 
    \log \mathcal{L} _{\rm{GR} }  (\theta) =c'-\frac{1} {2} \Gamma_{00}  \alpha^2_{\rm{ML} }  -\frac{1} {2}  (\theta -\theta_{\rm{ML} } ) ^i \Gamma_{ij}  (\theta -\theta_{\rm{ML} } ) ^j,
\end{equation} 

\begin{equation} 
    Z_{\rm GR} =\int d\theta \mathcal{L} _{\rm GR}  (\theta) =e^{c'-\Gamma_{00}  \alpha^2_{\rm{ML} } /2} 
    \sqrt{\frac{ (2\pi) ^k} {\det  \Gamma_{ij} } },
\end{equation} 
with  
\begin{subequations} 
    \begin{gather} 
        \alpha_{\rm ML} =\alpha_{\rm stat} +\alpha_{\rm sys}  \label{eq:alpha ML},\\
        \alpha_{\rm stat} =\frac{z} {\rho},\\
        \alpha_{\rm sys} =\sqrt{2\mathcal{M} }  \cos{\iota},  \\
        \sigma_\alpha=\frac{1} {\rho},\\
        \Gamma_{00} =\sigma^{-2} _\alpha \label{eq:gamma00},
    \end{gather} 
\end{subequations} 
where $z \sim N (0,1) $ is the random number associated with the noise realization, $\rho $ is the SNR of the signal, $\mathcal{M} $ is the mismatch given in Eq. \eqref{eq:mismatch}, $\sigma_\alpha$ is the standard deviation of the distribution, and $\iota $ is the angle between the signals $\Delta h (\theta_{\rm Tr} ) $ and $\partial h / \partial \alpha$.
The Bayes factor in favor of the deviation from GR is 
\begin{equation} 
  \mathcal{B} =\frac{\Pi} {A}   \frac{Z_{\rm non-GR} } {Z_{\rm GR} },
\end{equation} 
where $\Pi$ is the prior Bayes factor and $A=\alpha_{\rm max} -\alpha_{\rm min} $ is the prior range of $\alpha$.
With the assumption that $\Gamma_{0i} =0$, $\det \Gamma_{\mu \nu } =\Gamma_{00}  \det \Gamma_{ij} $, the Bayes factor simplifies to 
\begin{equation} \label{eq:B} 
  \mathcal{B} =\frac{\Pi} {A}  \sqrt{\frac{2\pi} {\Gamma_{00} } }  \exp (\frac{1} {2} \Gamma_{00} \alpha^2_{\rm ML} ).
\end{equation} 
Inserting the expressions in Eqs.  \eqref{eq:alpha ML}--\eqref{eq:gamma00}  into Eq. \eqref{eq:B}, the logarithm of the Bayes factor is given by 
\begin{equation} 
    \log \mathcal{B} =\log (\frac{\Pi} {A}  \frac{\sqrt{2\pi} } {\rho} ) +\frac{ (z+\rho \sqrt{2\mathcal{M} } \cos{\iota} ) ^2} {2}.
\end{equation} 
In this work,  our results are scaled to $\Pi=A=1$, $z=0$, $\cos{\iota}  \rightarrow {\cos{\iota} } _m=1$, and  the threshold of the Bayes factor $\mathcal{B} _{\rm threshold } =e^{10} $ as in Ref.   \cite{moore2021testing}. The final expression of $\log\mathcal{B} $ is 
\begin{equation} 
    \log \mathcal{B} =\log (\frac{\sqrt{2\pi} } {\rho} ) +\mathcal{M}  \rho^2.
\end{equation}

\begin{figure} 
\includegraphics[scale=0.8]{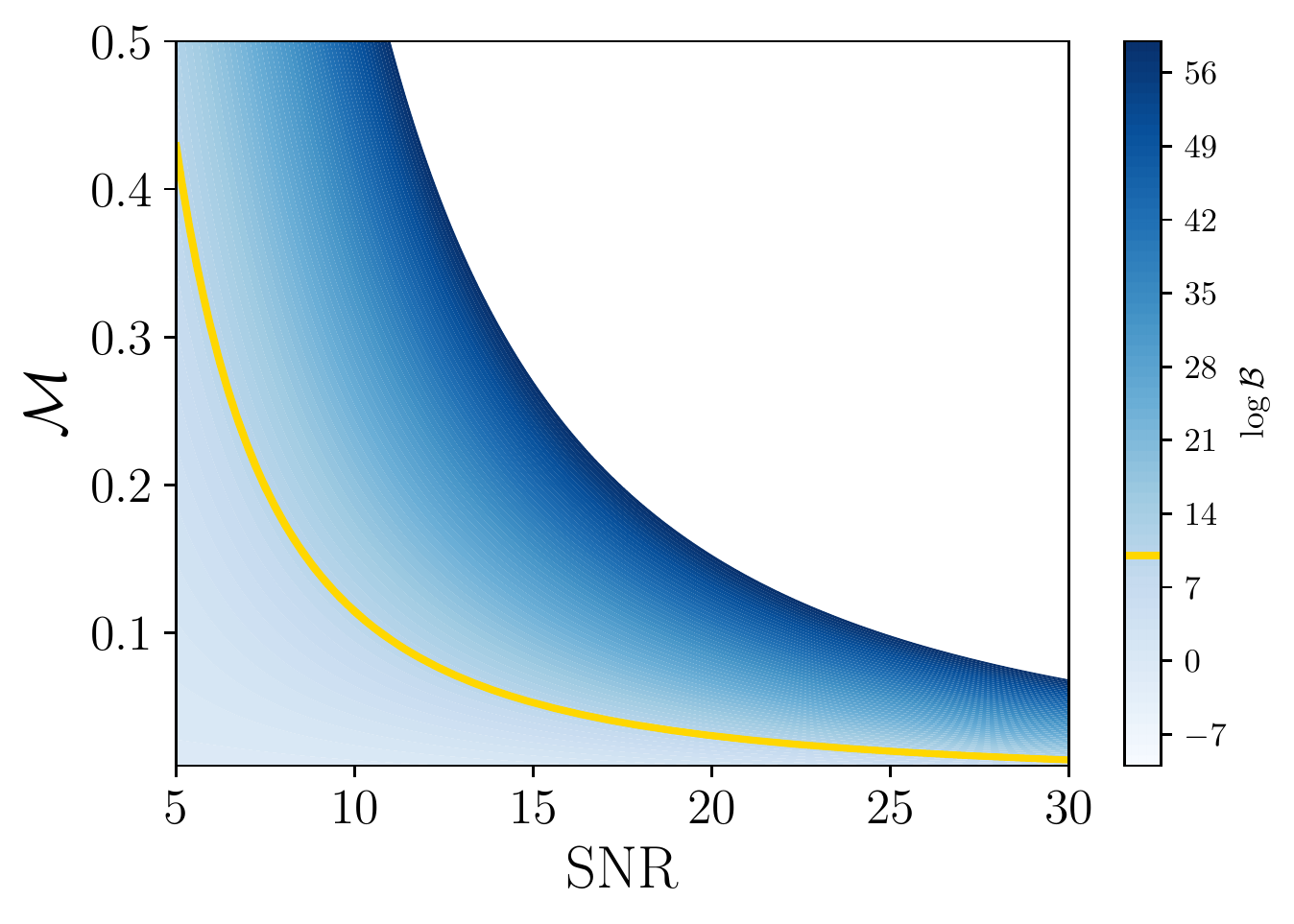} % Here is how to import EPS art
\caption{\label{fig:log B LARGE}  The logarithm of the Bayes factor, $\log\mathcal{B} $, of the  EMRI systems with different SNR and mismatch ($\mathcal{M} $)  values. The yellow line represents the threshold ($10$)  of $\log\mathcal{B} $.}  
\end{figure} 

Fig. \ref{fig:log B LARGE}  shows the logarithm of the Bayes factor as a function of SNR and mismatch ($\mathcal{M} $). We take the mismatches ranging from $10^{-2} $ to $0.5$ and SNRs ranging from $5$ to $30$. The yellow line represents the threshold ($10$) of $\log\mathcal{B}$, above which we may mistakenly claim to have detected a deviation from GR. We can see that even if the mismatch ($\mathcal{M} $)  is small, the calculated $\log \mathcal{B} $  may still exceed the threshold when the SNR is very large, which may induce a fake signal of deviation from  GR.

\begin{figure} 
\includegraphics[scale=0.7]{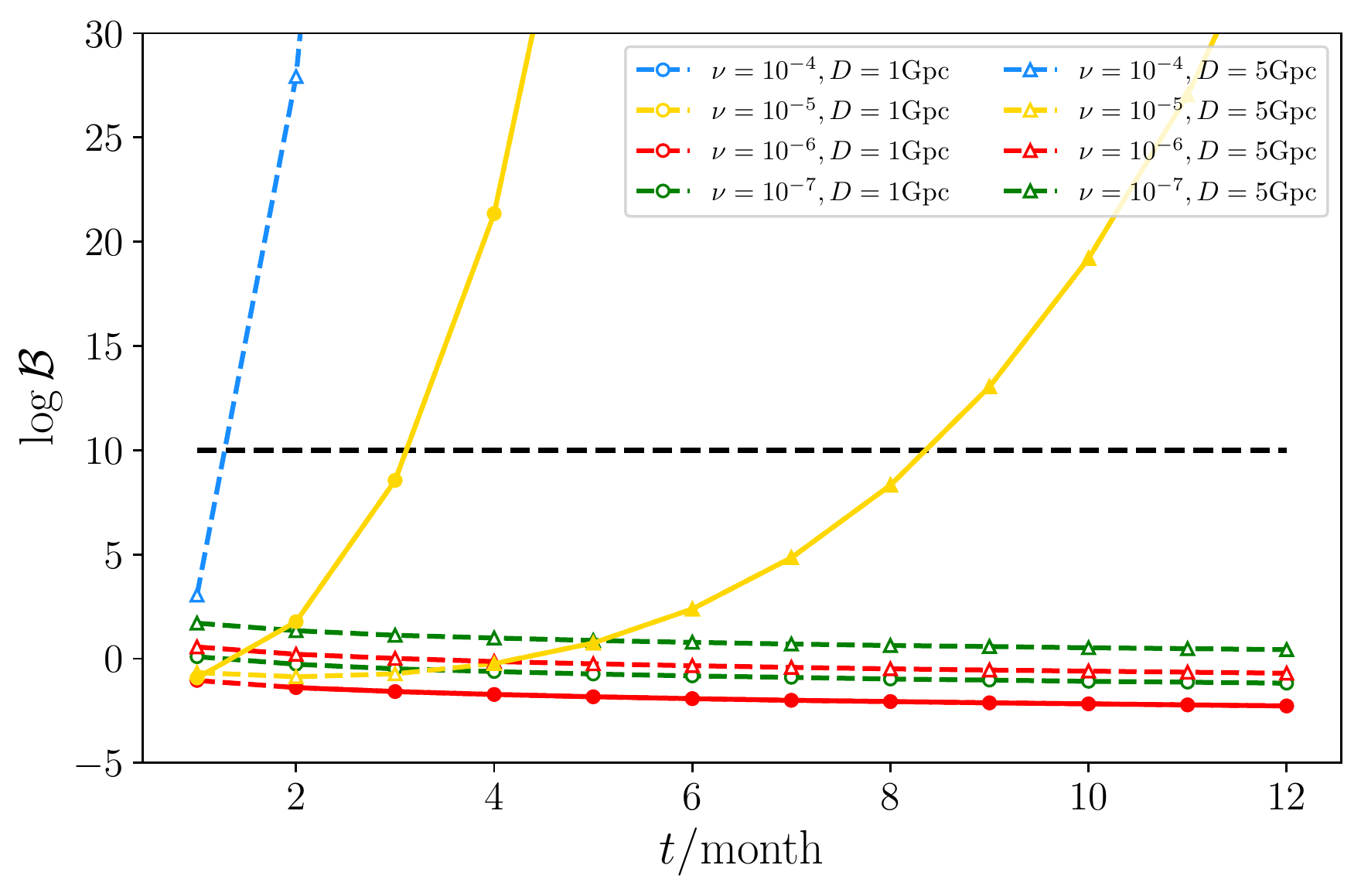} % Here is how to import EPS art
\caption{\label{fig:month compared}  The logarithm of the Bayes factor ($\log \mathcal{B} $)  versus evolution times for the  EMRI systems with the same orbital parameters in Fig. \ref{fig:mismatch}.  The dashed blue, yellow, red, and green lines represent systems with different mass-ratios of $10^{-4},10^{-5},10^{-6} $ and $ 10^{-7} $, respectively. The dashed black line denotes the threshold ($10$) of $\log \mathcal{B} $. The circles and  the triangles mark EMRIs at the distance of 1Gpc and 5Gpc, respectively. Solid triangles or circles connected by solid lines indicate EMRI systems with detectable SNRs greater than 10.} 
\end{figure}

As we described in Fig. \ref{fig:mismatch}, the mismatch of waveforms is related to the duration of signals. With the accumulation of time, the mismatch between the two signals  will increase. In addition, the SNR depends on both the luminosity distance of the source and the signal length.
As an example, Fig. \ref{fig:month compared}  shows the logarithm of the Bayes factor for different 
luminosity distances. We choose the same orbital parameters as in Fig. \ref{fig:mismatch}. Solid 
triangles or circles connected by solid lines indicate EMRI systems with detectable SNRs greater than 10.
Our results show that for the system with a mass-ratio of $\nu=10^{-4} $,  the calculated Bayesian factor will still exceed the threshold ($10$)  even when it is far away from us ($5$ Gpc)  due to its high mismatch and SNR, which may lead to a misjudgment of testing GR. For the case of $\nu=10^{-5} $, the source at 5 Gpc can also lead to the same misjudgment.
However, for the system with a mass-ratio of $\nu=10^{-6 \sim -7} $, due to its small 
mismatch, even if the SNR reaches a detectable threshold, it will not induce a 
false signal that deviates from GR predictions.

In Table.\ref{tab:logB}, for quantitative comparison, we list the logarithm of the Bayes factor ($\log \mathcal{B} $) due to mass-ratio for different twenty orbits.
We find that in the case of $\nu \gtrsim 10^{-5} $, when the SNR is high enough, the ignoring of the mass-ratio may indicate a deviation from the GR. However, for the EMRI system with a smaller mass-ratio, there is almost no possibility that we will make a wrong judgment.

\begin{table} [b]
\caption{\label{tab:logB} 
 The logarithm of the Bayes factors ($\log \mathcal{B} $) between two waveforms including or excluding mass-ratio corrections. 
 The signals continue for one year and end at $ (e_{\rm fin},\iota_{\rm fin},p_{\rm fin} ) $.} 

\begin{ruledtabular} 
\begin{tabular} {ccccccccccc} 
N & a    & $e_{\rm fin} $  &  $\iota_{\rm fin}  (^\circ) $ &  $p_{\rm fin} $ (M)   &    $\mu$  ($M_\odot$)   & $\nu$  & D (Gpc)  & $\mathcal{M} $ & SNR & $\log \mathcal{B}  $  \\
\hline
1  & 0.30     &  0.12      &  45    & 5.7    &   50    &  $ 1\times 10^{-4} $   &   2.5    & $ 8.4\times 10^{-1} $  &   150.1   & 18921.1 \\
2  &  0.43    &   0.10     &  72    &  7.2   &   20   &  $ 6\times 10^{-5} $   &   1.5   & $ 4.7\times 10^{-1} $    &  78.4    & 2885.4 \\

3  &  0.73   &   0.08     &  81    &  5.8   &   25   &  $ 6\times 10^{-5} $   &   7.0   & $ 7.4\times 10^{-1} $    &  16.2   &192.3 \\

4  &   0.14   &  0.16      &  70    & 6.9   & 35  &   $ 5 \times 10^{-5} $     &  2.0  &  $ 3.0\times 10^{-1} $   & 81.8   & 2003.9     \\ 
5  &  0.80    &   0.15     &   60   &  6.5  & 40  &    $ 4\times 10^{-5} $    & 6.0  &   $ 4.0\times 10^{-1} $   &  35.8 &  510.0   \\
6 &  0.35    &   0.13     &   35   &  6.4  & 30  &    $ 4\times 10^{-5} $    & 5.0  &   $ 1.9\times 10^{-1} $   &  51.8 &  506.8   \\
7  &  0.60   &   0.10     & 45    & 5.2    &  25    &   $ 3\times 10^{-5} $   & 3.0  &   $ 5.6 \times 10^{-1} $    & 62.6  & 2193.3  \\

8  &  0.70    &  0.20       &  10     &  6.0    &  20  &  $ 3\times 10^{-5} $   &  2.0 &  $ 4.4\times 10^{-1} $   &  80.7 &  2862.0 \\ 

9  &  0.80    & 0.10       &  70    &   5.2  &   40    &   $ 2\times 10^{-5} $    &     1.0  &   $ 3.4\times 10^{-1} $ & 134.4     &6137.6  \\ 

10  &   0.90   &0.20        & 85     &  6.3   &  45      &  $ 1\times 10^{-5} $     &    8.0  &   $ 2.4\times 10^{-2} $  &   9.0   &0.7 \\

11  &   0.70   &0.12        & 50     &  6.0   &  35      &  $ 1\times 10^{-5} $     &    1.5  &   $ 1.6 \times 10^{-2} $  &   78.5    & 95.2 \\

12  &   0.10   &  0.30      &  40    &  7.1   & 30  &   $ 8\times 10^{-6} $     &  3.0 &  $ 2.4\times 10^{-3} $   &  38.5 &  0.8   \\ 
13  &  0.60    &  0.22      &  45    &  7.2   & 40   &  $ 6\times 10^{-6} $  & 4.5   & $2.4 \times 10^{-4} $ &  24.6  & -2.1  \\

14  &  0.55    &  0.40      &  38    &  6.8   & 30   &  $ 5\times 10^{-6} $  &1.0   & $ 1.2\times 10^{-3} $ &  87.9 &   5.7\\ 
15  & 0.85     &  0.15      &   30    & 5.5     &  50   &  $ 4\times 10^{-6} $    & 2.0  &   $ 1.1\times 10^{-3} $   &   48.7     & -0.4   \\

16  & 0.20     &  0.35      &   55    & 7.0     &  30   &  $ 3\times 10^{-6} $    & 8.0  &   $ 2.5\times 10^{-4} $   &    7.1 &  -1.0\\

17  &   0.40    & 0.30        &  50    &  6.5   &  25    &  $ 2\times 10^{-6} $    & 5.0     &   $ 1.3\times 10^{-4} $   & 8.8    &    -1.2           \\

18  &  0.50   &  0.22      &  20    &  6.0   & 10   &  $ 2\times 10^{-6} $  & 4.0   & $5.0\times 10^{-4} $ &  8.9 &   -1.2\\

19   &  0.30     &   0.40      & 35      &  6.8    &    25  &   $ 1\times 10^{-6} $ & 1.0  &  $ 1.9\times 10^{-5} $  & 58.8   & -3.1  \\

20  &  0.62    &   0.30     &  60    & 5.8    &  10    &   $ 1\times 10^{-6} $   &  3.5  &   $  6.2\times 10^{-4} $    & 4.9  &  -0.7 \\

\end{tabular} 
\end{ruledtabular} 
\end{table}

\section{\label{conclusion} Conclusions} 

In this work, based on the EOB orbits and Teukolsky equation, we investigated the influence of mass-ratio corrections on the orbital motion (orbit, frequency, and phase)  and gravitational radiation. Our results show that even in the case of extreme mass-ratio, the mass-ratio  corrections may have a great impact on orbital frequency, and then affect orbital evolution. In addition,  we also illustrated the discrepancy in Teukolsky-based  energy fluxes caused by mass-ratio, in which case the gravitational radiation was considered. We find that for the relativistic orbits,  the relative difference of energy fluxes between the EOB orbit and the  test particle one is about 10$\nu$ magnitude, which can not be ignored. Both the difference in conservation dynamics and that in Teukolsky-based energy fluxes  may generate  accumulated errors during the long-term evolution of EMRIs and then affect the waveforms. This means that when constructing waveform templates of EMRIs, the mass-ratio cannot be ignored in the orbital calculation. In other words, if we neglect the mass-ratio corrections when building the EMRIs waveform templates, there is a risk of inducing a fake signal of deviation from GR.

One of the scientific goals of the study of EMRIs is to test GR and the nature of BHs. In order to support this goal, an accurate and effective waveform template is needed. However, this is still a challenge at present. The waveform template given in this work, combining EOB orbits and the Teukolsky equation, including mass-ratio corrections, is more accurate than the test particle one. Based on the Bayesian analysis, we evaluated the possibility of inducing a fake signal of deviation from GR because of omitting the mass-ratio corrections in waveform templates. We find that for the case of $\nu\lesssim 10^{-5} $, the mismatch  between the EOB and the  test particle orbits is so small  that  can be ignored. However, for the case of $\nu \gtrsim 10^{-5} $,  the mismatch is much bigger, and there is a risk of making the incorrect judgment that we have detected a deviation from GR. We may conclude that the mass-ratio correction in the EMRI waveform templates is important, especially when $\nu \gtrsim 10^{-5} $.
It is noted that we just use EOB dynamics to include the conservative mass-ratio correction which may not exactly describe the self-force effect at the extreme mass-ratio limit. However, our result should be kept at least qualitatively correct. We will compare our findings with existing waveforms such as those obtained from the GSF  formalism \cite{Barack_2018} or the numerical relativity (NR) simulations 
to validate our results and assess agreement between different approaches in the future. In our current study, we have not considered the spin of the secondary and its potential impact on the metric of the central black hole. However, in future work, we plan to investigate the effects of  the spin of the smaller object and estimate the order of magnitude of the spin contributions in our analyses \cite{PhysRevD.105.084031,PhysRevD.104.084067,Battista_2022}. Doing so could provide a more accurate picture of the dynamics of compact object inspirals in such extreme gravitational environments.

In our Bayesian parameter inference, we used a simple linearized analysis with the help of Fisher matrix calculation. 
We use a Gaussian distribution as a prior to describe the detector noise. It is known that this is not the case, due to a lot of transient noise that affects the sensitivity of the detector, so a more suitable prior could be used to improve our work.
We should point out that GR is assumed to be the correct description of nature in this work, which is not necessarily correct. If instead we use gravitational waveforms for data analysis but GR is wrong (stealth bias)   \cite{PhysRevD.89.022002}, what will happen to the results? 
A joint analysis will be conducted by using the waveforms with the GR assumption in the present paper and the non-GR one to confirm whether the ignoring of mass-ratio in waveform templates will induce misjudgment on testing GR in the forthcoming work. % that we have detected a deviation from GR.
In addition, here we only discussed the fake deviation of  GR caused due to the omitting of mass-ratio for a single GW event.
However, waveform errors may accumulate as the catalog size increases and lead to incorrect scientific conclusions   \cite{moore2021testing,Hu_2023}.
In the future, 
%we propose to enrich this work by using the bank of event catalogs.
we will also study the case of GW events in the catalog using a hierarchical Bayesian approach with the assumption that the non-GR parameters follow Gaussian distribution   \cite{PhysRevLett.123.121101}. 
We hope our work will be useful for developing EMRI waveform templates for space-borne  GW detectors.

\section{Acknowledgements*} 

This work was supported by the National Key R\&D Program of China (Grant Nos. 2021YFC2203002), the National Natural Science Foundation of China (Grant Nos. 12173071). Wen-Biao Han was supported by the CAS Project for Young Scientists in Basic Research (Grant No. YSBR-006). This work made use of the High Performance Computing Resource in the Core Facility for Advanced Research Computing at Shanghai Astronomical Observatory. We also thank Belahcene Imene for her valuable advice on this work.

\nocite{*} 

\appendix

\section{EOB formalism } 
  The metric potentials $A$ and $D$ for the EOB formalism mentioned in Ref.   \cite{steinhoff2016dynamical}  are given by
\begin{subequations} 
    \begin{gather} 
        A (u) =\Delta_u-\frac{a^2} {M^2} u^2 \\
        D^{-1}  (U)   = 1+\log[D_{\rm Taylor} ]        
    \end{gather} 
\end{subequations} 
with  
\begin{subequations} 
    \begin{gather}  %也可以用align + & 对齐
    \Delta_u=\Bar{\Delta} _u (\Delta_0 \nu +\nu \log (\Delta_5 u^5+\Delta_4 u^4+\Delta_3 u^3+\Delta_2 u^2+\Delta_1 u+1) +1) \\  
     \Bar{\Delta} _u  =\frac{a^2 u^2} {M^2}  +\frac{1} { (K\nu-1) ^2}   +\frac{2u} {K\nu-1}  \\    
   \Delta_5  = (K\nu-1) ^2[\frac{64} {5}  \log (u) + (-\frac{1} {3} a^2 (\Delta_1^3-3\Delta_1\Delta_2+3\Delta_3)  \\ \nonumber
      -\frac{\Delta_1^5-5\Delta_1^3\Delta_2+5\Delta_1^2 \Delta_3+5\Delta_1\Delta_2^2-5\Delta_2\Delta_3-  5\Delta_4\Delta_1} {5 (K\nu-1) ^2}  \\ \nonumber
      +\frac{\Delta_1^4-4\Delta_1^2\Delta_2+4\Delta_1\Delta_3+2\Delta_2^2-4\Delta_4} {2 (K\nu-1) }  \\ \nonumber
      +\frac{2275\pi^2} {512} +\frac{128\gamma} {5} -\frac{4237} {60} +\frac{256 \log (2) } {5} ]    \\  
     \Delta_4  =\frac{1} {96} [8 (6a^2 (\Delta_1^2-2\Delta_2)  (K\nu-1) ^2+3\Delta_1^4+\Delta_1^3 (8-8K\nu) -12\Delta_1^2\Delta_2+12\Delta_1 (2\Delta_2K\nu-2\Delta_2+\Delta_3)  \\ \nonumber
      +48\Delta_2^2-64 (K\nu-1)  (3\Delta_3-47K\nu+47) 
     -123\pi^2 (K\nu-1) ^2]   \\
     \Delta_3  =-a^2 \Delta_1 (K\nu -1) ^2-\frac{\Delta_1^3} {3} +\Delta_1^2 (K\nu -1) 
     +\Delta_1\Delta_2 -2 (K\nu -1)  (\Delta_2-K\nu +1)  \\
     \Delta_2  =\frac{1} {2}  (\Delta_1 (\Delta_1-4K\nu +4) -2a^2\Delta_0 (K\nu -1) ^2)  \\
     \Delta_1  = -2 (\Delta_0 +K)  (K\nu-1)  \\
     \Delta_0 = K (K\nu-2)    \\
     D_{\rm Taylor}  =1+6\nu u^2 +2 \nu u^3 (26-3 \nu) 
    \end{gather} 
\end{subequations} 
  
The Hamilton-Jacobi equation  is given by    \cite{damour2001coalescence} 
\begin{equation} \label{eq:H-J equation} 
    g^{\alpha \beta} P_\alpha P_\beta +\frac{Q_4 M^2 P^4_r} {r^2 \mu^2} +\mu^2=0
\end{equation} 
where the non-geodesic term $Q_4=\frac{2 (4-3\nu) \nu M^2} {r^2} $ appears at 3PN order and here we omit this term as in the Ref.   \cite{zhang2021geometrized}.
$P_t=-H_{\rm eff} $, $P_\phi=L_z$.
By substituting the metric expressions \eqref{eq:metric}  into Equation \eqref{eq:H-J equation}, we get four conserved quantities of motion

\begin{equation} \label{eq:Q} 
    \hat{Q} =\cos^2{\theta_{\rm min} } [a^2 (1-\hat{H} ^2_{\rm eff} ) +\frac{\hat{L} ^2_z} {\sin^2{\theta_{\rm min} } } ]
\end{equation} 
\begin{equation} \label{eq:Heff} 
    H_{\rm eff} =\frac{g^{t\phi} } {g^{tt} } P_{\phi}  + \frac{1} {\sqrt{-g^{tt} } } \sqrt{\mu^2+[g^{\phi \phi } -\frac{ (g^{t \phi} ) ^2} {g^{tt} } ]{P_{\phi} } ^2+g^{rr} P^2_r+g^{\theta \theta} P^2_{\theta} } 
\end{equation} 
\begin{equation} \label{eq:Lz} 
   \hat{L} ^2_z=\frac{ (a_1-a_2) ^2 (b_1+b_2) ^2- (b_1^2-b_2^2)  (b_1^2 c_1-b_2^2 c_2) 
   -2 (a_1-a_2) b_1 b_2 \sqrt{ (a_1-a_2) ^2- (b_1^2-b_2^2)  (c_1-c_2) } } 
   {[ (a_1-a_2) ^2- (b_1^2 c_1-b_2^2 c_2) ]^2}  
\end{equation} 
\begin{equation} \label{eq:E} 
    \frac{E^2} {M^2} =1+2 \nu (a_1 \hat{L} _z +\sqrt{ \frac{c_1\hat{L} ^2_z+1 } {b_1}  } -1) 
\end{equation} 
with 
\begin{subequations} 
    \begin{gather} 
        a_1=\frac{\Tilde{\omega} _{\rm fd1} } {\Lambda_{t1} }  \\
        a_2=\frac{\Tilde{\omega} _{\rm fd2} } {\Lambda_{t2} }  \\
        b_1=\sqrt{\frac{\Sigma_1\Delta_{t1} } {\Lambda_{t1} } }  \\
        b_2=\sqrt{\frac{\Sigma_2\Delta_{t2} } {\Lambda_{t2} } }  \\
        c_1=\frac{\Sigma_1} { (1-\cos^2{\theta_{\rm min} } ) \Lambda_{t1} }  \\
        c_2=\frac{\Sigma_2} { (1-\cos^2{\theta_{\rm min} } ) \Lambda_{t2} }  \\
    \end{gather} 
\end{subequations} 

With the help of these constants of motion, we  can drive the expressions of angular momentum by
\begin{subequations} \label{eq:angular momentum} 
    \begin{gather} 
        \hat{P_\theta} ^2=\Hat{Q} - \cos^2{\theta_{\rm min} } [a^2 (1-\hat{H} ^2_{\rm eff} ) +\frac{\hat{L} ^2_z} {\sin^2{\theta_{\rm min} } } ] \label{eq:ptheta} \\
        \hat{P_r} ^2=\frac{[a\hat{L} _z- (r^2+a^2) H^2_{\rm eff} ]^2- (r^2A (u) +a^2) [
        r^2+K+2\frac{\Tilde{\omega} ^2_{\rm fd} +ar^2 (A (u) -1) } {r^2A (u) +a^2} \hat{H} _{\rm eff} \hat{L} _z-G (r) \hat{L} ^2_z
        ]} { (r^2A (u) +a^2) ^2 D^{-1}  (u) }  \\
        P_\phi=L_z \label{eq:pphi}
    \end{gather} 
\end{subequations} 
where $P_\theta$, $P_r$, and $P_\phi$ are the polar, radial, and  azimuthal angular momentum.

From Eqs.  \eqref{eq:metric}, \eqref{eq:Q}--\eqref{eq:E}  and \eqref{eq:angular momentum}, the orbital evolution equations for $r$, $\phi$, $\theta$ can be obtained by
\begin{subequations} 
    \begin{gather} 
        \Dot{r} =\frac{\partial E} {\partial P_r} =-\frac{g^{rr} \hat{P_r} } {E/M (g^{tt}  \hat{H} _{\rm eff} -g^{t \phi}  \hat{L} _z)  } \\
        \Dot{\phi} =\frac{\partial E} {\partial P_\phi} =\frac{g^{t \phi} -[g^{tt} g^{\phi \phi} - (g^{t \phi} ) ^2] \frac{\hat{L} z} {g^{tt}  \hat{H} _{\rm eff} -g^{t \phi}  \hat{L} _z} } {g^{tt} E/M} \\
        \Dot{\theta} =\frac{\partial E} {\partial P_\theta} =-\frac{g^{\theta \theta} \hat{P_\theta} } {E/M (g^{tt}  \hat{H} _{\rm eff} -g^{t \phi}  \hat{L} _z) } 
    \end{gather} 
\end{subequations} 
\iffalse
we define $\xi$ and $\chi$ by the following equations 
\begin{equation} \label{eq:chi} 
    \cos^2{\theta} =\cos^2{\theta_{\rm min} }  \cos^2{\chi} =\sin^2{\iota}  \cos^2{\chi} 
\end{equation} 
\begin{equation} \label{eq:zeta} 
    r=pM/ (1+e\cos{\xi} ) 
\end{equation} 
where $\iota$ is the orbital inclination, $\chi$ varies from $0$ to $2\pi$, and $\xi$ varies from $0$ to $2\pi$.
\fi
The  parameters in the expressions of the coordinate-time frequencies $\omega_{r} $, $\omega_{\theta} $, $\omega_{\phi} $ (Eqs.  \eqref{eq:eob-omega-r}--\eqref{eq:eob-omega-phi} )  are expressed by
\begin{subequations} 
    \begin{gather} 
         K (k) =\int_{0} ^{\pi/2}  \frac{d \chi} {\sqrt{1-k \sin^2{\chi} } } \label{eq:K} \\
         E (k) =\int_{0} ^{\pi/2}  \sqrt{1-k \sin^2{\chi} }  d \chi \\
         \Pi (z^2_- ,k) =\int_{0} ^{\pi/2}  \frac{d \chi} { (1-z^2_- \sin^2{\chi} ) \sqrt{1-k \sin^2{\chi} } }  \\
         X=\int_{0} ^{\pi}\frac{ep\sin{\xi}}{(1+e\cos{\xi})^2} \frac{d\xi}{\Delta_r P_r}\\
         z_-=\cos{\theta_{\rm min} }  \\
         z_+=\frac{L^2_z+Q+\beta +\sqrt{ (L^2_z+Q+\beta) ^2-4\beta Q} } {2 \beta}  \\
         \beta^2=a^2 (\nu^2-H_{\rm eff} ^2) \\
         k= z^2_-/ z^2_+ \label{eq:k}
    \end{gather} 
\end{subequations} 

\section{2PN fluxes } 
The hybrid scheme of fluxes proposed by  Gair and Glampedakis   \cite{gair2006improved}  are given by
\begin{equation} \label{eq:Lz mod} 
\begin{split} 
    (\Dot{L}_z ) _{\rm mod}  = (1-e^2) ^{3/2}  [ (1-e^2) ^{-3/2}  (\Dot{L}_z ) _{\rm 2PN}  (p,\iota, e, a)  - (\Dot{L}_z ) _{\rm 2PN}  (p,\iota, 0, a) + (\Dot{L}_z ) _{\rm fit} ]
\end{split} 
\end{equation} 

\begin{equation} \label{eq:Q mod} 
\begin{split} 
    (\Dot{Q} ) _{\rm mod}  =& (1-e^2) ^{3/2}  \sqrt{Q (p,\iota, e, a) } [ (1-e^2) ^{-3/2}  (\frac{\Dot{Q} } {\sqrt{Q} } ) _{\rm 2PN}  (p,\iota, e, a)  - (\frac{\Dot{Q} } {\sqrt{Q} } ) _{\rm 2PN}  (p,\iota, 0, a)  \\
    & +2\tan \iota ( (\Dot{L}_z ) _{\rm fit} +\frac{ \sqrt{Q (p,\iota, 0, a) } } {\sin^2{\iota} }  (\Dot{\iota} ) _{\rm fit} ) ]
\end{split} 
\end{equation} 

\begin{equation} \label{eq:E mod} 
\begin{split} 
     \Dot{E}  =& (1-e^2) ^{3/2}  [ (1-e^2) ^{-3/2}  (\Dot{E} ) _{\rm 2PN}  (p,\iota, e, a)  - (\Dot{E} ) _{\rm 2PN}  (p,\iota, 0, a)  
     -\frac{N_4 (p,\iota) } {N_1 (p,\iota) }  (\Dot{L}_z ) _{\rm fit}  (p,\iota, 0, a)  \\
     &-\frac{N_5 (p,\iota) } {N_1 (p,\iota) }  (\Dot{Q} ) _{\rm mod}  (p,\iota, 0, a) ]
\end{split} 
\end{equation} 
with
\begin{equation} \label{eq:Edot2pn} 
\begin{split} 
     (\Dot{E} ) _{\rm 2PN}  = &-\frac{32} {5}  \frac{\mu^2} {M^2}  (\frac{M} {p} ) ^5 (1-e^2) ^{3/2} [
   (1 +\frac{73 e^2} {24} + \frac{37 e^4} {96} ) 
      -q (\frac{M} {p} ) ^{3/2}  (\frac{73} {12} +\frac{823 } {24} e^2+\frac{949 } {32} e^4+\frac{491 } {192} e^6) \cos{\iota} \\
     &- (\frac{M} {p} )  (\frac{1247} {336} +\frac{9181} {672} e^2) +\pi (\frac{M} {p} ) ^{3/2}  (4+\frac{1375} {48} ) 
    - (\frac{M} {p} ) ^{2}  (\frac{44711} {9072} +\frac{172157} {2592} e^2) \\
    &+q^2 (\frac{M} {p} ) ^{2}  (\frac{33} {16} +\frac{359} {32} e^2) 
    -\frac{527} {96} q^2 (\frac{M} {p} ) ^{2} \sin^2{\iota}  
          ]       
\end{split} 
\end{equation} 

\begin{equation} \label{eq:Lzdot2pn} 
\begin{split} 
     (\Dot{L}_z ) _{\rm 2PN}  = &-\frac{32} {5}  \frac{\mu^2} {M}  (\frac{M} {p} ) ^{7/2}  (1-e^2) ^{3/2}  [
    (1+\frac{7} {8} e^2) \cos{\iota} + q (\frac{M} {p} ) ^{3/2} \{ (\frac{61} {24} +\frac{63} {8} e^2+\frac{95} {64} e^4) \\
    &-\cos^2{\iota}  (\frac{61} {8} +\frac{91} {4} e^2+\frac{461} {64} e^4) \} 
     - (\frac{M} {p} )  (\frac{1247} {336} +\frac{425} {36} e^2) \cos{\iota} 
    +\pi (\frac{M} {p} ) ^{3/2}  (4+\frac{97} {8} e^2) \cos{\iota}  \\
    &- (\frac{M} {p} ) ^{2}  (\frac{44711} {9072} +\frac{302893} {6048} e^2) \cos{\iota} 
     +q^2 (\frac{M} {p} ) ^{2} \cos{\iota}  ( (\frac{33} {16} +\frac{95} {16} e^2) -\frac{45} {8} \sin^2{\iota} ) 
     ]
\end{split} 
\end{equation} 

\begin{equation} \label{eq:Qdot} 
\begin{split} 
     \Dot{Q}  =& -\frac{64} {5} \frac{\mu^2} {M}  (\frac{M} {p} ) ^{7/2}  \sqrt{Q} \sin{\iota}  (1-e^2) ^{3/2} [
      (1+\frac{7} {8} e^2) -q (\frac{M} {p} ) ^{3/2} \cos{\iota}  (\frac{61} {8} +\frac{91} {4} e^2+\frac{461} {64} e^4) \\
      & - (\frac{M} {p} )  (\frac{1247} {336} +\frac{425} {36} e^2) 
    +\pi (\frac{M} {p} ) ^{3/2}  (4+\frac{97} {8} e^2) 
    - (\frac{M} {p} ) ^{2}  (\frac{44711} {9072} +\frac{302893} {6048} e^2) \\
    & +q^2 (\frac{M} {p} ) ^{2}  ( (\frac{33} {16} +\frac{95} {16} e^2) -\frac{45} {8} \sin^2{\iota} ) 
     ]
\end{split} 
\end{equation} 

\begin{equation}  \label{eq:Lzfit}
\begin{split}
     (\Dot{L}_z ) _{\rm fit}  =&-\frac{32} {5}  \frac{\mu^2} {M}  (\frac{M} {p} ) ^{7/2} [
    \cos{\iota} +q  (\frac{M} {p} ) ^{3/2}  (\frac{61} {24} -\frac{61} {8} \cos^2{\iota} ) -\frac{1247} {336}  (\frac{M} {p} ) \cos{\iota} \\
    &+4\pi (\frac{M} {p} ) ^{3/2}  \cos{\iota} -\frac{44711} {9072}  (\frac{M} {p} ) ^{2}  \cos{\iota} 
     +q^2 (\frac{M} {p} ) ^{2}  \cos{\iota}  (\frac{33} {16} -\frac{45} {8} \sin^2{\iota} ) \\
     &+ (\frac{M} {p} ) ^{5/2} \{
     q (d_1^a+d_1^b (\frac{M} {p} ) ^{1/2} +d_1^c (\frac{M} {p} ) ) 
     +q^3 (d_2^a+d_2^b (\frac{M} {p} ) ^{1/2} +d_2^c (\frac{M} {p} ) ) \\
     &+   \cos{\iota}  (c_1^a+c_1^b (\frac{M} {p} ) ^{1/2} +c_1^c (\frac{M} {p} ) ) 
    +q^2\cos{\iota}  (c_2^a+c_2^b (\frac{M} {p} ) ^{1/2} +c_2^c (\frac{M} {p} ) ) \\
     &+q^4\cos{\iota}  (c_3^a+c_3^b (\frac{M} {p} ) ^{1/2} +c_3^c (\frac{M} {p} ) ) 
    +q\cos^2{\iota}  (c_4^a+c_4^b (\frac{M} {p} ) ^{1/2} +c_4^c (\frac{M} {p} ) ) \\
  &+q^3\cos^2{\iota}  (c_5^a+c_5^b (\frac{M} {p} ) ^{1/2} +c_5^c (\frac{M} {p} ) ) 
   +q^2\cos^3{\iota}  (c_6^a+c_6^b (\frac{M} {p} ) ^{1/2} +c_6^c (\frac{M} {p} ) ) \\
   &+q^4\cos^3{\iota}  (c_7^a+c_7^b (\frac{M} {p} ) ^{1/2} +c_7^c (\frac{M} {p} ) ) 
  +q^3\cos^4{\iota}  (c_8^a+c_8^b (\frac{M} {p} ) ^{1/2} +c_8^c (\frac{M} {p} ) ) \\
   &+q^4\cos^5{\iota}  (c_9^a+c_9^b (\frac{M} {p} ) ^{1/2} +c_9^c (\frac{M} {p} ) ) 
        \}  \\
        &+ (\frac{M} {p} ) ^{7/2} q \cos{\iota} \{
                         (f_1^a+f_1^b (\frac{M} {p} ) ^{1/2} ) 
        +               q (f_2^a+f_2^b (\frac{M} {p} ) ^{1/2} ) 
        +             q^2 (f_3^a+f_3^b (\frac{M} {p} ) ^{1/2} ) \\
        &+   \cos^2{\iota}  (f_4^a+f_4^b (\frac{M} {p} ) ^{1/2} ) 
        +  q\cos^2{\iota}  (f_5^a+f_5^b (\frac{M} {p} ) ^{1/2} ) 
        +q^2\cos^2{\iota}  (f_6^a+f_6^b (\frac{M} {p} ) ^{1/2} ) 
        \} 
     ]
\end{split} 
\end{equation} 

\begin{equation}  \label{eq:iotafit}
\begin{split} 
     (\Dot{\iota} ) _{\rm fit}  =& \frac{32} {5}  \frac{\mu^2} {M} q\frac{\sin^2{\iota} } {\sqrt{Q} }  (\frac{M} {p} ) ^{5/2} [
     \frac{61} {24} + (\frac{M} {p} )  (d_1^a+d_1^b (\frac{M} {p} ) ^{1/2} +d_1^c (\frac{M} {p} ) ) 
     +q^2 (\frac{M} {p} )  (d_2^a+d_2^b (\frac{M} {p} ) ^{1/2} +d_2^c (\frac{M} {p} ) )  \\
     & +q \cos{\iota}  (\frac{M} {p} ) ^{1/2}  (c_{10} ^a+c_{10} ^b (\frac{M} {p} ) ^{1/2} +c_{10} ^c (\frac{M} {p} ) ) 
    +q^2 \cos^2{\iota}  (\frac{M} {p} )  (c_{11} ^a+c_{11} ^b (\frac{M} {p} ) ^{1/2} +c_{11} ^c (\frac{M} {p} ) ) \\
    & + (\frac{M} {p} ) ^{5/2}  q^3 \cos{\iota}  \{
     (f_7^a+f_7^b (\frac{M} {p} ) ^{1/2} ) 
        +               q (f_8^a+f_8^b (\frac{M} {p} ) ^{1/2} ) \\
        &+   \cos^2{\iota}  (f_9^a+f_9^b (\frac{M} {p} ) ^{1/2} ) 
        +  q\cos^2{\iota}  (f_{10} ^a+f_{10} ^b (\frac{M} {p} ) ^{1/2} ) 
            \} 
     ]
\end{split} 
\end{equation} 
where 
\begin{equation} 
\begin{split} 
     &d_1^a=-10.7420, d_1^b=28.5942,d_1^c=-9.07738,
     d_2^a=-1.42836    , d_2^b= 10.7003  ,d_2^c= -33.7090  ,\\
     &c_1^a=-28.1517    , c_1^b= 60.9607  ,c_1^c=40.9998   ,
     c_2^a=-0.348161    , c_2^b=2.37258   ,c_2^c=-66.6584   ,\\
     &c_3^a=-0.715392    , c_3^b= 3.21593  ,c_3^c=5.28888  ,
     c_4^a=-7.61034    , c_4^b=128.878   ,c_4^c=-475.465   ,\\
     &c_5^a=12.2908    , c_5^b=-113.125   ,c_5^c=306.119   ,
     c_6^a=  40.9259  , c_6^b=-347.271   ,c_6^c=886.503   ,\\
     &c_7^a=-25.4831    , c_7^b=224.227   ,c_7^c= 490.982  ,
     c_8^a= -9.00634   , c_8^b=91.1767   ,c_8^c=-297.002   ,\\
     &c_9^a=-0.645000    , c_9^b= -5.13592  ,c_9^c= 47.1982  ,
     c_{10} ^a=-0.0309341    , c_{10} ^b=-22.2416   ,c_{10} ^c=7.55265   ,\\
     &c_{11} ^a=-3.33476    , c_{11} ^b=22.7013   ,c_{11} ^c=-12.4700   ,
     f_1^a= -283.955   , f_1^b= 736.209, 
     f_2^a=483.266    ,\\
     &f_2^b= -1325.19   ,
     f_3^a=-219.224    , f_3^b=634.499     ,
     f_4^a= -25.8203   , f_4^b= 82.0780   ,
     f_5^a=301.478    ,\\ 
     &f_5^b= -904.161 ,
     f_6^a=-271.966    , f_6^b=  827.319   ,
     f_7^a=-162.268    , f_7^b=  247.168 ,
     f_8^a=  152.125  ,\\
     &f_8^b=  -182.165,  
     f_9^a=184.465    , f_9^b=-267.553,   
     f_{10} ^a=-188.132    , f_{10} ^b=254.067 .  
\end{split} 
\end{equation} 

 However, the mass-ratio correction is not involved in the expressions of their 2PN fluxes. In this work, we make some corresponding improvements  by adding mass-ratio correction terms in  Eqs.  \eqref{eq:Edot2pn}--\eqref{eq:Qdot}  to get the evolution expressions of orbit parameters $ (p,e,\iota) $.  Our improved 2PN fluxes are given by \cite{Zhang_2021,PhysRevD.96.104048,10.1093/ptep/ptv092}

\begin{equation} \label{eq:newEdot2pn} 
\begin{split} 
     (\Dot{E} ) _{\rm 2PN}  = &-\frac{32} {5}  \frac{\mu^2} {M^2}  (\frac{M} {p} ) ^5 (1-e^2) ^{3/2} [
   (1 +\frac{73 e^2} {24} + \frac{37 e^4} {96} ) 
      -q (\frac{M} {p} ) ^{3/2}  (\frac{73} {12} +\frac{823 } {24} e^2+\frac{949 } {32} e^4+\frac{491 } {192} e^6) \cos{\iota} \\
     &- (\frac{M} {p} )  (\frac{1247} {336} +\frac{5\nu} {4} +e^2 (\frac{9181} {672} +\frac{325\nu} {24} ) -e^4 (\frac{809} {128} -\frac{435\nu} {32} )  -e^6 (\frac{8609} {5376} -\frac{185\nu} {192} )  ) \\
     &+\pi (\frac{M} {p} ) ^{3/2}  (4+\frac{1375} {48} ) 
    - (\frac{M} {p} ) ^{2}  (\frac{44711} {9072} +\frac{172157} {2592} e^2) 
    +q^2 (\frac{M} {p} ) ^{2}  (\frac{33} {16} +\frac{359} {32} e^2) 
    -\frac{527} {96} q^2 (\frac{M} {p} ) ^{2} \sin^2{\iota}  
          ]       
\end{split} 
\end{equation} 

\begin{equation} \label{eq:newLzdot2pn} 
\begin{split} 
     (\Dot{L} _z) _{\rm 2PN}  = &-\frac{32} {5}  \frac{\mu^2} {M}  (\frac{M} {p} ) ^{7/2}  (1-e^2) ^{3/2}  [
    (1+\frac{7} {8} e^2) \cos{\iota} + q (\frac{M} {p} ) ^{3/2} \{ (\frac{61} {24} +\frac{63} {8} e^2+\frac{95} {64} e^4) \\
    &-\cos^2{\iota}  (\frac{61} {8} +\frac{91} {4} e^2+\frac{461} {64} e^4) \} 
     - (\frac{M} {p} )  (\frac{1247} {336} +\frac{7\nu} {4} +e^2 (\frac{425} {36} +\frac{401\nu} {48} ) -e^4 (\frac{10751} {2688} -\frac{205\nu} {96} )     ) \cos{\iota} \\
    &+\pi (\frac{M} {p} ) ^{3/2}  (4+\frac{97} {8} e^2) \cos{\iota}  
    - (\frac{M} {p} ) ^{2}  (\frac{44711} {9072} +\frac{302893} {6048} e^2) \cos{\iota} 
     +q^2 (\frac{M} {p} ) ^{2} \cos{\iota}  ( (\frac{33} {16} +\frac{95} {16} e^2) -\frac{45} {8} \sin^2{\iota} ) 
     ]
\end{split} 
\end{equation}

\begin{equation} \label{eq:newQdot} 
\begin{split} 
     \Dot{Q}  =& -\frac{64} {5} \frac{\mu^2} {M}  (\frac{M} {p} ) ^{7/2}  \sqrt{Q} \sin{\iota}  (1-e^2) ^{3/2} [
      (1+\frac{7} {8} e^2) -q (\frac{M} {p} ) ^{3/2} \cos{\iota}  (\frac{61} {8} +\frac{91} {4} e^2+\frac{461} {64} e^4) \\
      & - (\frac{M} {p} )  (\frac{1247} {336} +\frac{7\nu} {4} +e^2 (\frac{425} {36} +\frac{401\nu} {48} ) -e^4 (\frac{10751} {2688} -\frac{205\nu} {96} )    ) 
    +\pi (\frac{M} {p} ) ^{3/2}  (4+\frac{97} {8} e^2) 
    - (\frac{M} {p} ) ^{2}  (\frac{44711} {9072} \\
    &+\frac{302893} {6048} e^2) 
     +q^2 (\frac{M} {p} ) ^{2}  ( (\frac{33} {16} +\frac{95} {16} e^2) -\frac{45} {8} \sin^2{\iota} ) 
     ]
\end{split} 
\end{equation} 

Our final hybrid scheme of fluxes is given by Eqs. \eqref{eq:Lz mod}--\eqref{eq:E mod}  with Eqs. \eqref{eq:Lzfit}--\eqref{eq:iotafit} and Eqs. \eqref{eq:newEdot2pn}--\eqref{eq:newQdot}.

\bibliography{eobgw} 

\end{document}